# Understanding the Perceived Relevance of Capability Measures: A Survey of Agile Software Development Practitioners

Sai Datta Vishnubhotla[1], Emilia Mendes and Lars Lundberg
Department of computer science,
BTH - Blekinge Tekniska Högskola, Sweden
[1]Corresponding author. E-mail address: sai-datta.vishnubhotla@bth.se

## ABSTRACT

**Context:** In the light of the swift and iterative nature of Agile Software Development (ASD) practices, establishing deeper insights into capability measurement within the context of team formation is crucial, as the capability of individuals and teams can affect team performance and productivity. Although a former Systematic Literature Review (SLR) synthesized the state of the art in relation to capability measurement in ASD – with a focus on selecting individuals to agile teams, and capabilities related to team performance, productivity and success, determining to what degree the SLR's results apply to practice can provide progressive insights to both research and practice.

**Objective:** Our study investigates how agile practitioners perceive the relevance of individual and team level measures for characterizing the capability of an agile team and its members. Here, the emphasis was also on selecting individuals to agile teams, and capabilities associated with effective teams in terms of their performance, productivity and success. Furthermore, to scrutinize variations in practitioners' perceptions, our study further analyzes perceptions across stratified demographic groups.

**Method:** We undertook a Web-based survey using a questionnaire built based on the capability measures identified from a previously conducted SLR.

**Results:** Our survey responses (60) indicate that 127 individual and 28 team capability measures were considered as relevant by the majority of practitioners. We also identified seven individual and one team capability measure that have not been previously characterized by our SLR. The surveyed practitioners suggested that an agile team member's *responsibility* and *questioning skills* significantly represent the member's capability.

**Conclusion:** Results from our survey align with our SLR's findings. Measures associated with social aspects were observed to be dominant compared to technical and innovative aspects. Our results can support agile practitioners in their team composition decisions.

*Keywords:* individual capability; team capability; capability measurement; agile team formation; survey; agile software development

# 1. Introduction

Human aspects in the field of software development are considered as fundamental constituents that ultimately give a project team its soul [1]. A recent study investigating human aspects within software engineering [2] reported that a major portion of previous research and practice have focused on exploring technological or process related aspects, thereby leaving behind the social and psychological aspects sparsely explored. With the rapid rate of adoption of Agile Software Development (ASD) methods across various organizations over the past decades (e.g., [3], [4], [5], [6]), the necessity for collaborative work requiring multidisciplinary skills and interpersonal skills has increased [7].

Within ASD, a failure in assigning a suitable person to a team can ultimately lead to a decline in service quality and unattainable project deadlines [8], [9]. In order to provide adequate mechanisms for reacting to changing markets and reducing lead time, ASD methodologies are largely reliant on the capabilities of professionals [10]. These capabilities pertain to human aspects such as the qualities, features, social and methodological abilities that are key for professional and personal development [11], [12]. The capabilities of a professional influence team outcomes [13], and team capability is a crucial factor that leads the path towards high team performance, productivity, successful teams, and ultimately project success [14]. Therefore, considering individual and team capabilities while assigning professionals to various roles is crucial for establishing effective ASD teams [15].

In recent times, research studies in ASD emphasized on human aspects while investigating aspects such as team composition [16], [17], characteristics of high-performance teams [18], [19], and personality profiles of software engineers [20], [21]. The majority of such studies, however, discussed capabilities only to some extent and devoted very little attention towards identifying a comprehensive set of skills and abilities concerning individuals or teams. This was clearly evident from the findings of our previous study, a Systematic Literature Review (SLR) [22] conducted to synthesize the state of the art relating to attributes and criteria used for measuring individual and team capability in ASD, within the context of team building criteria, individual/team performance, productivity, and successful teams.

Despite the use of a rigorous search and selection processes in our previously conducted SLR, only two of the sixteen primary studies targeted exclusively at capabilities of agile practitioners. The other fourteen studies were looking more broadly at factors affecting individual/team productivity, performance and success in ASD, and were only included in our SLR because, amongst the identified factors, there were capability factors that could be related to selecting people for teams. Such difference in focus, in addition to a limited number of research studies exploring capabilities in ASD, may pose a threat to the external validity of the SLR findings with regard to ASD practice.

One of the mechanisms of ensuring the degree of scientific value of an SLR's findings, to both research and practice, is to rigorously assess its validity [23]. An obvious way to do so, as also accomplished in previous Software Engineering (SE) studies (e.g., [24], [25], [26]), is by collaborating with practitioners and determining to what degree they find the SLR's results relevant [27].

While a study with high research relevance has a higher potential for utility in industry and could subsequently increase the potential of research impact [27], it was reported that a major portion of SE studies possessed high methodological rigor but yet failed to address the important issue of relevance [27]. So, the driving question motivating the current study is to determine whether, and to what extent, the state of the art, in relation to capability measurement in ASD, is applicable to practice. To the best of our knowledge, none of the former SE studies analyzed agile practitioners' perception of measures for characterizing the capability of an agile team and its members.

The novelty of our current study is twofold: first, finding a meaningful intersection between the state of the art and practice by means of analyzing what measures (results from our former SLR) do practitioners perceive as relevant for characterizing capability in ASD, when targeting at team formation, team performance, productivity and success. Second, uncovering how the perceived relevance of capability measures vary among groups of practitioners, by means of a detailed contextual and statistical analysis. In this regard, a survey was conducted by recruiting professionals from diverse domains, where they rated the relevance of individual and team level measures for characterizing the capability of an agile team and its members. In the light of no clear and established standards for identifying survey subjects over social media, this study discussed and employed a search-string-based strategy for recruiting practitioners possessing a specific set of skills or competencies. The survey responses were analyzed to find dominant capability measures and to further study the differences in the perceptions of practitioners.

Investigating the degree to which capability measures apply to practice, would provide insights to software organizations for making decisions regarding human resource management [11]. The capability measures validated by diverse agile practitioners can be used by managers to assess the socio-technical skills of members and make decisions while assigning members to roles and allocating members to teams. Additionally, the assessment of capabilities can help managers in identifying the need for training, collaborating activities and designing counter measures, which can subsequently lead to strengthening the abilities and skills of personnel and teams [28].

This investigation also serves as an intermediate study for our ongoing research collaboration with two software development organizations in Sweden (a large sized telecom company [company A] and a small sized software company that offers services and products in mobile data services [company B]). The findings from this study, i.e. the set of measures that were vetted by agile practitioners as highly relevant for characterizing capability, would further be used towards forecasting agile team climate at company A [28] and towards developing an agile team performance forecasting model at company B [29].

With regard to the contribution and implication to research, this is the first study to bring in the perspectives from state of the art and practice in relation to capability measurement within ASD, when focusing upon team formation, productivity, performance and success. The results from our descriptive analysis clearly show that measures in relation to social aspects outweigh those linked to professional and innovative aspects while characterizing capability of individuals and teams. Further, the additional capability measures indicated by some of the respondents complement our SLR findings. The differences in practitioners' perceived relevance of certain capability measures, especially distinctions within groups of agile practitioners, indicate that other factors could also potentially be influencing their judgement, and thus, open venues for further research.

The remaining of this paper is structured as follows: Section 2 describes related work on capability measurement in ASD, Section 3 presents the methodological details of the study and demographic details of respondents. Section 4 reports the results of statistical analyses. In Section 5, we compare our key findings with existing studies and present the limitations of this study. Further, the implications and lessons learned, the threats to validity of this study and comments on future work are also presented in Section 5. Finally, Section 6 states the conclusions from our study.

# 2. Related work

Former literature in SE includes a significant number of studies that investigated a variety of human aspects like competencies of team members (e.g., [9], [30], [31], [32]), as well as, soft skills, social skills and personalities (e.g. [33], [34], [35]). Due to the prevalence of ASD methods and the need for personal and interpersonal skills while working in teams [15], [22], within ASD, there has been an appreciable progress in the research involving human aspects [36], [37].

Capability relates to the qualities and features that can be used or developed by individuals and teams [12], [22]. Additionally, the capability definitions reported in various studies [38], [39] convey that capability also pertains to the factors influencing success, software productivity and performance. Based on the capability definitions, our former SLR [22] retrieved various measures from the SE literature that characterize the capability of individuals and teams in ASD, when targeting at team building criteria, individual/team performance, productivity and successful teams. These capability measures pertain to professional, social and innovative aspects such as knowledge, skills, personality characteristics, abilities, aptitudes and attitudes.

Within SE, the findings from the studies where the voices of practitioners are acknowledged, would make a significant contribution to both research and practice. In this regard, we have attempted to identify the set of studies that discussed different capability aspects, within industrial agile contexts, by gathering and analyzing practitioners' opinions. We started by inspecting the primary studies of our SLR and found five relevant studies (studies in Table 1 without an *). Next, we used the forward-snowballing technique over Google Scholar to search for the citations to our SLR, and also the citations to each of the five relevant primary studies. The snowballing technique was iterated until no more relevant studies could be found. This led to the identification of five relevant studies from recent years (i.e., studies published after our SLR search process was concluded).

In order to identify relevant studies published after our SLR concluded, we have executed another search iteration over four online databases (Scopus, Science direct, ACM and Wiley. Note: These were the four databases used primarily in the SLR's search process) using the same search string from our SLR. However, this search did not retrieve any new potential studies beyond the five studies identified by snowballing. Overall, the list of studies which explored capability aspects by gathering practitioners' opinions are presented in Table 1, and among these, the five recent studies have been highlighted with a '*' mark. The details of the studies in Table 1 are presented as follows:

Multiple studies reported skills in relation to individuals [29], [40], [41], [42], [43]. These studies emphasized on aspects like non-technical skills, soft skills, factors influencing tacit knowledge transfer. Among these, the majority of the studies (five) were executed within the context of Scrum methodology [29], [40], [41], [42], [43]. The five studies commonly used interviews to gather data and among them, commitment [29], [40], customer orientation [40], [43] and communication skills [42], [43] were reported by multiple studies as valued skills.

Matturro et al. [40] examined the 'insider' voices of Scrum practitioners, i.e. about the soft skills they consider most valued to have by Scrum master and product owner roles. The authors interviewed 25 experienced Scrum practitioners from eight companies and identified communication skills and teamwork to be most valued by both the roles. Besides them, customer orientation was identified to be valuable for program managers and commitment, responsibility, interpersonal and planning skills were considered to be important for Scrum masters.

Melo et al. [41] determined the factors that impact an agile team's productivity by gathering the opinions of software engineers who were part of teams that used agile methods (XP or Scrum) for at least two years. By means of semi-structured interviews (19 professionals), followed by observations, face-to-face discussions and retrospective documentation review, the authors identified personnel factors like personality, full-time allocation and knowledge levels as key factors to be considered while aligning agile teams.

Takpuie and Tanner [42] examined the factors impacting tacit knowledge transfer in Scrum teams. The authors organized 12 semi-structured interviews with Scrum team members at two companies. Using thematic analysis, the authors identified the characteristics that interviewees opined to be crucial among the team members who were able to successfully transfer knowledge. This study reported personnel factors like motivation, credibility, empathy and communication skills as key for knowledge transfer in Scrum teams.

Matturro et al. [43] studied how the role of product owner is performed in industrial practice, to present similarities and differences between the literature and what is observed in the field. The authors conducted semi-structured interviews with six product owners who were part of four different companies. The semi-structured interviews targeted at gathering the product owners' perceptions of aspects such as relationship with team members, and most valued skills in a product owner. All the interviews were transcribed and analyzed using open coding. Results showed that communication skills, teamwork and customer orientation are the three soft skills that a good product owner should exhibit.

In Vishnubhotla et al. [29], which was one of our previous studies, we used interviews and grounded theory to investigate the capabilities and criteria used by managers (two senior professionals) while assembling Scrum teams in a small sized software organization. Apart from project specific factors and organizational factors, our study identified individual capability measures such as developers' domain knowledge, developers' own interest, previous deliverables' quality to be crucial while building teams.

On the other hand, four studies reported team level skills and abilities, within ASD context [24], [44], [45], [46]. While two studies emphasized on identifying team level factors influencing agile team's productivity [24], [44], the rest focused on developing a team tacit knowledge measure [45] and investigating critical success factors in ASD projects [46]. The studies employed interviews [44], survey [46] and a mix of both mechanisms [24], [45], for gathering data. Upon comparing the team level factors discussed among studies, we noticed team's motivation was reported by all the four studies as an important characteristic.

Melo et al. [44] investigated team level factors influencing an agile team's productivity. The authors interviewed 13 agile team members and performed thematic analysis to determine the factors influencing productivity. The factors like team experience, competencies, motivation and communication were among the list of various factors identified to be impacting an agile team's productivity.

Ryan and O'Connor [45] developed and validated a team tacit knowledge measure for managing agile teams. The authors initially conducted unstructured interviews with 13 experts to identify the factors influencing team performance. Subsequently, to ascertain which of the constructs were true, they analyzed the opinions of novices and experts (18 professionals). Finally, the tacit knowledge measure was validated using 48 teams. This study identified team level factors like high motivation, cooperation, experience, morale, competition and clear goals as key factors.

Chow and Cao [46] conducted a survey for identifying the critical success factors of ASD projects. The authors gathered opinions of agile practitioners from 25 countries. They identified 12 possible success factors by analyzing the details of 109 projects reported by the respondents and consolidated the factors into five categories. In terms of the impact of the five categories on agile project success, the people dimension (motivation and expertise) was reported to be most important.

Fatema and Sakib [24] explored agile team members' perception of productivity influencing factors. They used a system dynamics approach and employed interviews (12 professionals) and survey (17 professionals) to gather the perceptions of agile practitioners, in the context of Bangladesh software Industry. The four most perceived factors identified in this study were team effectiveness, team management, motivation and customer satisfaction.

**Table 1. Studies discussing capability aspects within industrial agile contexts based on gathering practitioners' opinions**

| Target entity | Study | Research method(s) and data collection mechanism(s) | Research topic |
|---|---|---|---|
| Individual engineers | [40] | Field- study: interviews | Identifying soft skills valued by Scrum practitioners |
| | [41] | Case-study: interviews, observations, face-to-face discussions, and retrospective documentation review | Identifying personnel factors influencing agile team's productivity |
| | [42]* | Interpretivist epistemology: interviews | Identifying factors impacting tacit knowledge transfer in Scrum teams |
| | [43]* | Semi-structured interviews | Identifying procedures and criteria used to select people for the role of product owner |
| | [29]* | Case study: interviews | Investigating criteria used for team building |
| Teams | [44] | Case-study: interviews | Identifying team level factors influencing agile team's productivity |
| | [45] | Mixed-method: interviews and survey | Developing and validating team tacit knowledge measure |
| | [46] | Survey | Investigating critical success factors in ASD projects |
| | [24]* | Mixed-method: literature review, interviews and survey | Identifying team level factors influencing agile team's productivity |
| Individual engineers and teams | [47]* | Case study: interviews | Investigating criteria used for team building |

* Studies published after SLR [22] was concluded.

We observed that almost all of the previous studies listed in Table 1 (except for [47]) examined only specific capability aspects in relation to either individual professionals or teams, but not both. There were studies that investigated professional aspects (e.g., [48], [40], [41]). One of our previous studies [47] was the only one that examined capability aspects from the perspective of both individuals and teams.

In Mendes et al. [47], which was one of our previous studies, a case-study was executed in a large size software company that practiced agile methods. We investigated the capabilities and criteria used by senior professionals (14 professionals) while allocating people to tasks and inquired about the factors affecting team performance. By means of interviews and grounded theory procedures, the study identified 10 individual and five team capability measures.

Unlike the aforementioned studies, our study herein investigates multiple dimensions of capabilities (professional, social and innovative dimensions), by gathering the perceptions of professionals associated with agile teams from diverse domains. We have identified only one previous study [11] that seems close to ours, as elaborated next.

Moustroufas et al. [11] reviewed the SE literature and presented a competency profiling model. This model listed competencies of a software engineer in relation to professional, social and innovative aspects. This model was not developed exclusively for supporting agile methodologies. However, it was tested within a telecom company that practiced ASD methods. The authors reported that professionals from the company perceived their model to be detailed and informative. However, no details were specified regarding the procedure used for gathering the perceptions of professionals.

In the wake of the significance of team members' capabilities in bringing agility to a development process [41] and the pressing need for assigning capable professionals to teams and projects [38], [39], identifying which capability measures are highly regarded by practitioners when considering team formation, performance, productivity and success, is a crucial topic for investigation. To the best of our knowledge, none of the previous SE studies investigated which individual and team level measures would be appropriate for characterizing the capability of an agile team and its members, while aiming at building effective agile teams.

# 3. Research method

By means of an SLR [22], we already gathered evidence on the state of the art pertaining to capability measurement within ASD. Consequently, for the sake of exploring the state of the practice, (i.e., to understand which capability measures pertain to software professionals' practice), a methodology that elicits the knowledge of agile practitioners has to be employed. Therefore, we adopted survey method for this study as it is a well-established means for gathering information regarding the experience and expertise affiliated with a reasonably well-defined community [49].

In this study, our interest lies in understanding what practitioners' perceived relevance of various capability measures is, rather than exploring why. So, we chose to conduct a descriptive survey as opposed to: a) explanatory survey, which is used for investigating the reasons for occurrence of a phenomenon and b) exploratory survey, which is used as a pre-study to look for patterns and ideas before testing any hypotheses [50], [51].

We acknowledge that organizing interviews would have given us two-fold benefits, firstly, interviews would have given us the flexibility of posing targeted questions to interviewees, which help in gaining an in-depth opinion. Secondly, interviews would have permitted us to explain any misunderstandings in questions [50]. However, as we planned to execute a descriptive survey and were interested in collecting information from a large sample, we decided to use a questionnaire as data collection mechanism. One advantage of using questionnaire is that, our team of researchers and respondents need not synchronize time and place for data collection and this point enables us to recruit participants across the world [50].

Planning a survey involves tackling multiple challenges pertaining to its design. The event of a faulty survey design not only poses a threat to the validity of the results but also limits the scope for its replication [49]. So, in order to reduce researcher bias while conducting our survey and to ensure its rigor and repeatability, we initiated this investigation by preparing a protocol.

The discussions by Kitchenham and Pfleeger [51] on how surveys can be used to address SE topics and their suggestions and recommendations on the issues that need to be considered while using a survey, were incorporated in the design of our study. In the light of the existence of multiple guidelines that emphasize on different aspects of SE surveys (e.g., [52], [53]), we used an empirically evaluated checklist to guide our survey design and audit our survey report [54]. This comprehensive checklist was designed by systematically aggregating knowledge from 12 methodological studies and it focused on reviewing aspects of evidence and reporting. The key aspects of the protocol together with the details of how those aspects were executed will be discussed in the rest of this section.

## 3.1. Research questions

Upon inspecting the details of all the studies listed in Table 1, we can notice that there has been no comprehensive multi-aspect study that focused exclusively on reporting the views of agile practitioners from various organizations, regarding the qualities, features, skills and abilities concerning individuals and teams associated with ASD. Moreover, none of the former studies attempted to bring in the perspectives from both research and practice towards identifying a meaningful intersection of capability measures. So, we have formulated our first research question to address this gap and this question aims to determine what measures practitioners perceive as relevant for representing capability in ASD, while considering team formation, productivity, performance and success.

Some of the studies conducted in other contexts reported that perceptions of software professionals differ based on their role [40], work experience [55] and work environment (ASD methodology [56] and team size [57]). However, a closer inspection of the studies reported in Table 1 reveals that none has analyzed the differences in the perceptions of different demographic groups of practitioners. In the light of an empirically evaluated checklist for surveys [54], which recommends the analysis of data according to stratified demographic groups for the sake of drawing meaningful comparisons, we have formulated our second research question; this is another research contributions of this paper. The second research question aims to determine how the perceived relevance of capability measures vary among different groups of practitioners. The demographic dimensions asked in the questionnaire were used to create different sub-groups (sub-groups henceforth), which were employed for investigating variations in practitioners' perceptions.

With the goal of identifying a meaningful intersection of capability measures that would be relevant for both research and practice, our survey majorly includes closed-ended questions, framed in connection to capability measures from our SLR. But, in order to gather additional measures from respondents, which they see to be relevant, the respondents need to be given an opportunity and space for providing freestyle answers i.e., to report additional measures in their own words and in as much detail as they like. Therefore, our survey also includes two open-ended questions to gather additional measures that characterize individual and team capability. We have formulated the third research question to determine the additional measures from the open-ended questions. In essence, the research questions guiding this study are:

- RQ.1) What measures do practitioners perceive as relevant for representing capability in ASD?
  - RQ.1.1) What is the perceived relevance of individual capability measures?
  - RQ.1.2) What is the perceived relevance of team capability measures?
- RQ.2) How does the perceived relevance of capability measures vary among different sub-groups?
  - RQ.2.1) What (if any) are the differences in the relevance levels of capability measures indicated by various sub-groups?
  - RQ.2.2) Which individual and team capability measures are perceived as more relevant among various sub-groups?
- RQ.3) In addition to the list of capability measures specified in the questionnaire, what are the other additional measures that agile practitioners think are relevant for characterizing individual and/or team capability?

Responses to these research questions provide the empirical foundation for identifying crucial capability measures within the context of ASD. This enables researchers and agile practitioners to understand which capability measures are deemed relevant for individuals, teams and also for a particular role and ASD methodology. By bringing in the perspectives from scientific literature and industrial practice, we believe that our study contributes towards better realizing the area of capability measurement in ASD.

## 3.2. Survey design

The comprehensive catalogues of capability measures that evolved from our former SLR [22] were instrumental in driving our survey. The results from our SLR illustrate both individual and team capability measures by primarily grouping them into professional, social and innovative categories. Then, under each of these three categories, the measures were further classified into pertinent sub-categories. The categorization of capability measures in our SLR was done using the same classification suggested in one of the primary studies [11]. Figure 1 presents an overview of how the findings from the SLR are classified. Here, the first capability measure can be seen to be primarily classified under the professional category and then subsequently classified into the software requirements sub-category. The catalogues of capability measures along with the descriptions and categorizations that have been used in our survey can be found in Appendix A of the supplementary material.

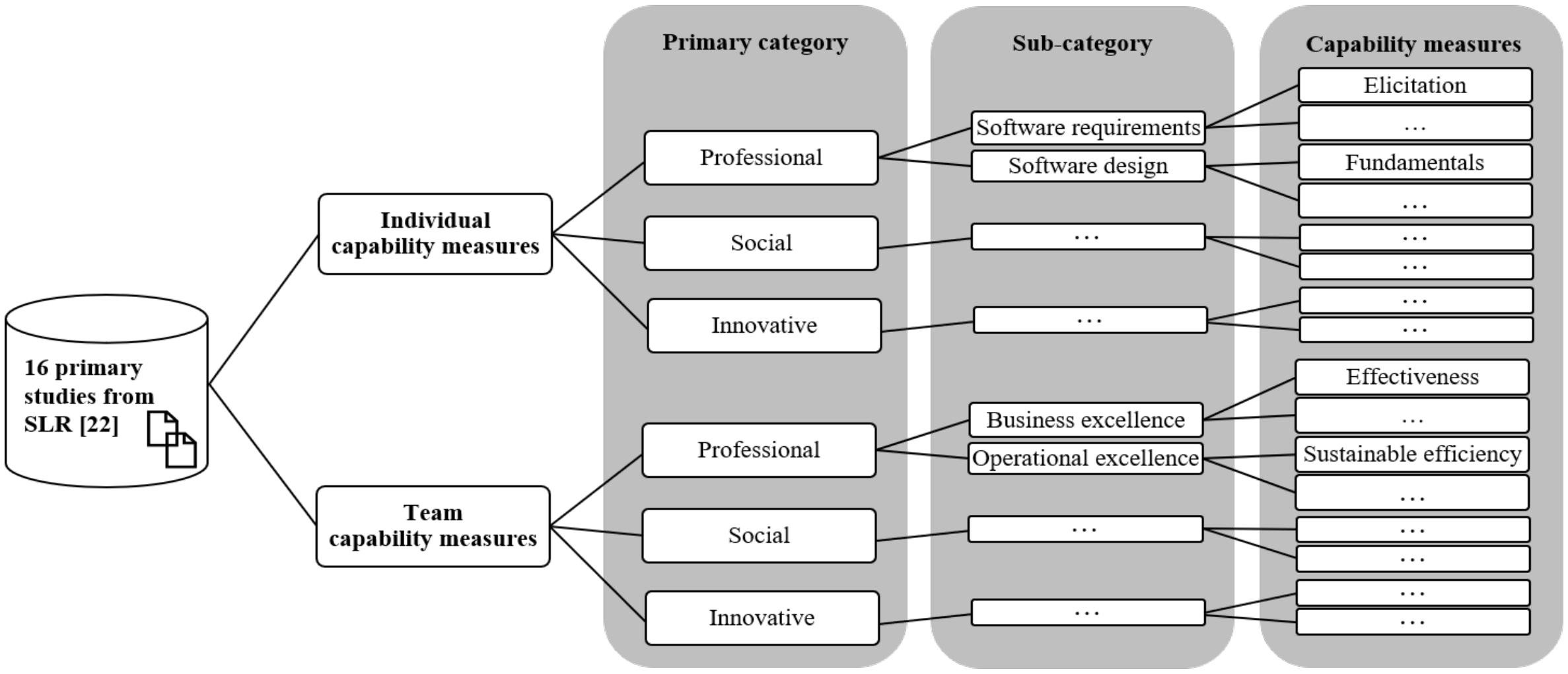

**Figure 1. Categorization of capability measures in SLR**

### 3.2.1. Population and sampling technique

Determining population is a key element in survey design and in our case, the target population was global community of software professionals possessing experience of working in ASD teams. In general, the sampling methods associated with survey based data collection are broadly classified into probabilistic and non-probabilistic sampling [51], [58]. Although conducting probabilistic sampling [51] would be ideal for the phase of finding subjects, in our case, due to the difficulty associated with finding an obvious forum for identifying and approaching a large number of professionals from different organizations[1], such a sampling was not practically possible. Thus, we chose to adopt a non-probabilistic sampling method, which has been reported as an acceptable approach [59].

The process of collecting non-probabilistic sampled data and then performing a systematic sampling [51] (to mimic the probabilistic sampling), i.e. pooling data from members belonging to various organizations and then selecting the data so that every member of the population has seen a statistically equal chance of being selected, would not be feasible in our case as we chose to record completely anonymous responses. This is because, in cases where organization names or any revealing information are recorded, participants might be hesitant to sharing their honest opinions, which, consequently has been reported to have serious impact on the quality of data reported [58], [60].

In cases where systematic sampling is not possible, techniques associated with non-systematic sampling can be applied [59]. Convenience sampling is one such technique and it is employed where it would be practically difficult to include individuals from a large population. Thereby, in convenience sampling, participants are selected based on researcher’s accessibility and hence, the method is considered to be the most common among all sampling techniques [58].

Although there exists a criticism that this technique introduces bias to the sample, a highly cited survey in SE [61] reported that convenience sampling is the dominant survey and experimental approach in SE. This sampling technique is also popularly used in other disciplines such as medicine research (e.g., [62], [63], [64]) and social sciences (e.g., [65], [66], [67]). So, despite the limitations and bias involved, reasons like limited access to ASD practitioners and maintaining anonymity among survey responses led us to the selection of convenience sampling for this study.

[1] Although we identified some relevant LinkedIn groups, they explicitly stated that their space emphasizes on promoting discussions among members and put forth restrictions on using the space for advertising surveys.

Additionally, we also used snowball sampling [51] for recruiting subjects. This technique has been observed to be particularly useful in hard-to-reach populations, where the researcher tries to contact a network of qualified study subjects with the support of first point contacts [68]. In our study, the people recruited during the convenience sampling phase were asked to nominate other people they believe to be suitable and willing to take part in our survey.

### 3.2.2. Recruitment of subjects

Subjects for the survey were recruited in two phases. In the first phase, people from personal contacts and those recommended by colleagues were approached. The personal contacts of the three authors of this study consisted of software practitioners having different levels of experience. Especially, the second and third authors' active engagement in multiple industry-academia research collaborations helped us in, both, identifying software organizations practicing agile methodologies and approaching practitioners from those organizations. Further, some of our former research studies executed in association with agile teams situated at our industrial collaborator's site, opened doors for us for contacting practitioners who were already aware of our research area.

Murphy et al. [69] mentioned that, by definition, survey research is a social interaction between researchers and potential respondents, in short, 'a conversation with a purpose'. They opined that the methods employed by individuals to carry out such conversations have changed overtime and for the sake of survey research, they suggested that a researcher should employ the tools that the targeted population currently use to carry out conversations.

LinkedIn has been listed as the most popular cross-industry professional network, which provides researchers a great opportunity for collecting data from experienced personnel [70]. The prime advantage of using it is the ability to apply filters, which permit a researcher to target prospective participants [71]. Although there are currently no widely established standards for using LinkedIn in identifying subjects [70], [71], and social-media in general, a representative sample can still be acquired when a researcher makes an initial contact and request for participation, with appropriate individuals belonging to the population of interest [70].

In the second phase of recruiting subjects, we used LinkedIn for searching and contacting potential subjects. On LinkedIn, subjects relevant to our study were identified by means of the self-listed skills that an individual's profile specifies to be proficient in. In order to identify people who were well-versed in agile methodologies, we executed a search on LinkedIn by means of a Boolean string that included multiple ASD methodology names along with the keyword 'agile' (e.g., skills = scrum OR dynamic systems development OR feature driven development OR agile) and then filtered out the first (immediate) and second-degree connections from the search results. People among the second-degree connections were approached with the help of mutual connections.

Our sampling frame acquired from the aforementioned two phases (prior to snowball sampling) consisted of a list of 350 people and it is important to note that we did not restrict the sampling frame with respect to the number of years of practical experience. Instead, we included individuals if they had experience with working in real-world projects. The effectiveness of using personal network for recruiting software practitioners has been reported to depend upon the quality and quantity of the network [72]. We believe the process of consolidating social media connections, professionals associated with our industrial collaborator and the ones recommended by our colleagues, make our personal network diverse and prodigious.

### 3.2.3. Survey execution

The means of administering a survey influences the organization of items in the survey questionnaire [49]. So, herein, we initially focus on describing the details of survey implementation. The survey was administered through formsite, a commercial website which provides secure access, tools for managing the survey and options for analyzing the responses. Besides these, the biggest benefit of this platform is the feasibility for building custom forms using HTML and CSS, an option which is not available over major survey development/hosting websites.

Exploring the range of graphical options available on the formsite platform helped us in determining the options that could be employed for recording responses to the survey questions. For example, we decided to provide radio buttons for the question inquiring primary role and for the question regarding ASD method(s) currently being implemented in one's team, we chose to display a list and provide a checkbox beside each methodology name. For both the questions, in case a respondent could not find any relevant answer, with a click on 'other' option, a text field was provided for entering other response. Further, we decided to display the description of each capability measure by means of tooltips.

## 3.3. Data preparation and collection

We developed a Web-based self-administered semi-structured questionnaire, which comprised a mix of structured and unstructured questions that were intended to gain deeper insights into the subjects' responses. While the questionnaire majorly used closed questions, making the survey faster and easier to complete, three questions in the last section of the questionnaire used an open format in order to seek explanatory information (one of the questions was for collecting feedback). The closed questions could be answered using Likert item or pre-defined set of answers [51]. Whereas open format allowed for numeric answers or free text. The complete questionnaire can be found in Appendix B of the supplementary material.

We have taken multiple measures to make sure that respondents interpreted the questions correctly. We included a brief paragraph in the questionnaire's start page stating the study's aim and describing the types of questions presented in each section. Furthermore, we presented the definitions of capability at the start of a section and made them visible while answering the questions. We further made sure to present all the questions in relation to capability measures as closed-ended questions, in order not to lead to different interpretations. To facilitate easy interpretation of each capability measure, we stated that the description of a capability measure could be seen upon hovering the mouse pointer over that measure. Further, it was explicitly mentioned that there were no right, or wrong answers and the respondents were asked to provide their immediate impressions.

Next, the details of how the survey responses will be stored and processed were presented. Here, it was mentioned that the responses would be kept anonymous and soon after the survey, all the responses would be downloaded and securely stored on a system which would not be accessible to any person except our research team. For maintaining anonymity of subjects, we decided not to include any questions seeking the subject's name, email ID or organization.

One of the most important ethical principles concerning human subject research is in relation to receiving a subject's full informed consent to participate in a research study [73]. So, the questionnaire's start page hosted an informed consent form to comply with ethical principles. This form stated that participation in the survey was voluntary. Access to the rest of the questionnaire was blocked until consent was given.

The first section of our questionnaire consisted of demographic questions for gathering information about respondents' work profile. These questions were important not only in terms of survey documentation and ensuring reliability [50] but, they were beneficial for filtering and grouping subjects, which was essential for answering RQ.2. In the light of the fact that we did not have control over the survey referrals done by our colleagues, some of the demographic questions were built into the survey design to confirm whether respondents belonged to the targeted population. The second and third sections of the questionnaire consisted of closed questions pertaining to individual and team capability measures, respectively. In order to help respondents interpret the notion of capabilities, the definitions of individual and team capability [22] were presented at the start of their respective sections. Then, the research questions presented in the Section 3.1 determined the structure for collecting data in these two sections.

Within the second and third sections, each question was formulated in relation to a sub-category (see Figure 1) highlighted in the catalogues of capability measures; and the question was presented as a matrix of measures and rating options i.e., for each of the capability measures falling under a sub-category, the respondents were asked to rate the relevance over a four-point Likert item (with categories: highly relevant, relevant, somewhat relevant and irrelevant) or to select the 'I don't know' option in case they were unaware of the capability measure.

The categories of the four-point Likert item were formerly used in the context of studying perceptions of subjects [74], [75]. We intended to record the impressions of the respondents by means of encouraging them to indicate positive or negative perspective over each measure and this was the reason behind using a four-point Likert item without any neutral category. Considering the fact that not all participants understand all technologies, the last option (I don't know') was included to deal with the diverse background of participants [76]. Appendix B of the supplementary material presents the complete details of the questionnaire's second (questions ranging from eight to 30) and third sections (questions ranging from 31 to 38).

While the subjects were given an opportunity to rate the measures in any order, all the questions in sections two and three were marked as mandatory. This was done as a precautionary measure to avoid the incidence of unanswered questions, which has been reported to be, unfortunately, a common phenomenon in online SE surveys [58], [59], [77]. Finally, the last section of the questionnaire included three open-ended questions in order to collect any other additional measures that respondents thought would be relevant and also provided space for entering feedback on the survey.

In a study discussing the best practices and guidelines for online data collection, King et al. [71] recommended that online studies should be brief in order to minimize participant burden and maximize data collection. They reported 200 questions, requiring no more than 40-45 minutes, as a ceiling for online questionnaires. Beyond that, subjects were observed to be giving up without answering all the questions. Based on these insights, in our questionnaire, we limited the total number of questions to 42, of which, 31 focused on gathering the relevance levels of 161 measures.

In order to make sure that the terminology in our survey questions was coherent and consistent, we consulted two software professionals prior to the pilot process and sought their advice on the descriptions of capability measures. Further, their feedback was collected on the initial draft of survey questions. Receiving practitioners' feedback during the design phase of a survey has been observed to be a common practice among survey studies (e.g., [58], [77], [78]) and in our case, the two professionals acknowledged that the terminology used in the questionnaire was familiar.

Next, the initial version of the questionnaire was hosted on the formsite platform. Before we began the actual data collection process, two external researchers who were a part of a different research project and who had prior industrial experience of working in ASD teams, piloted our survey. Among the two researchers, one of the researchers, who has a PhD, proposed guidelines for conducting surveys in SE [54], [79] and the other one is a certified Scrum master. Both researchers previously undertook various roles in agile teams. The unique expertise of these members, from both academic and industrial dimensions, helped us in testing the validity and readability of our questionnaire. We asked the researchers to inspect whether the language used in the questions was concrete and simple. This process was key in terms of ensuring the clarity of the survey and for receiving suggestions on the presentation of our questionnaire. The researchers completed the entire questionnaire in a single session, taking an average time of 40 minutes. They used the free-text space provided in the last section of the questionnaire for providing feedback. The researchers' feedback contributed to improving the questionnaire's layout. Their suggestions, such as highlighting capability category's name in questions and presenting a progress bar on the webpage for indicating how many questions were left to be answered, were incorporated into the final version of the questionnaire.

Once the survey questionnaire was ready for circulation, an email was composed with a brief text explaining the aim survey and attaching the Web-link to the survey. This email further requested people to share the survey with others who might be interested. The email was then sent to invite the potential subjects identified among personal networks. In the case of individuals identified over LinkedIn, the email contents were sent as a direct message, using InMail feature. An approach for improving response rate of surveys is to send follow-up mail to non-respondents after a suitable interval [49]. In our case, the non-respondents could not be traced as we chose to conduct an anonymous survey. So, we decided to send only one follow-up email approximately two weeks after sending the initial invitation.

## 3.4. Data analysis

We received 87 responses in total, of which 60 (69%) were complete. Among the 87 responses, in 21 cases, only demographic questions were answered and in four cases, the questionnaire was partly completed. The rest were responses from the pilot session. Upon inspecting the partly filled questionnaires, we observed that the few answered questions contribute very little to our study and therefore decided to exclude all of them from our analysis. The two responses from the pilot session were also not included in the data analysis as they were used for validation purposes. This resulted in considering a total of 60 completed questionnaires for the data analysis phase. These 60 usable responses corresponded to a response rate of 17% from the original sampling frame.

All the 60 responses were downloaded from the formsite platform as a consolidated MS excel sheet. Answer coding was applied where required (e.g., transforming Likert scale entries to integers for facilitating ordinal comparisons) prior to analyzing the respondents' entries. We have then exported this information to a .CSV file and used the R programming language and statistical software environment for our analysis. While the first author took care of curating the data, both first and second authors contributed towards the selection of tests and data analysis. The first author then took the lead in preparing the survey report and the rest reviewed results and provided feedback.

To analyze the survey data, we used descriptive statistics as well as scoring criteria and tests for studying the differences among stratified demographic groups. First, the frequencies of responses, in percentage, were calculated over the Likert item for each measure and this information was used to plot heatmaps across different categories of measures. These heatmaps help in visualizing the overall responses and highlight the measures that are predominantly regarded as highly relevant or irrelevant. Next, filtering and sorting criteria were applied to the frequency information to rank capability measures within a primary category, and also across the three primary categories. These ranks aid towards segregating a list of capability measures that were perceived as highly relevant (Section 4.1).

In relation to the goal of analyzing variations in the perceptions across practitioners, the Kruskal-Wallis H test was applied to investigate whether there were any differences in the ratings among demographic sub-groups (Section 4.2.1). Next, a scoring scheme was proposed to compare proportions of Likert responses within each sub-group. These scores were used to determine which measures were widely indicated as relevant within the sub-groups (Section 4.2.2).

Data from all the questionnaire items was used for answering the research questions, especially, the ordinal data from the second and third questionnaire sections was used for data analysis associated with RQ.1 and RQ.2. The demographic details from the first section of the questionnaire were further used to filter responses into various sub-groups, which were important for answering RQ.2. Finally, the attributes indicated via the free-text space provided in the last section of the questionnaire were used to answer RQ.3.

## 3.5. Survey demographics

When the subjects were asked to indicate the geographical location of their organization, a major portion of the subjects indicated their organization to be located in Sweden (85%), followed by India and Germany (5% each), Brazil, Finland and the United States (1.6% each). We observed that the subjects were associated with diverse software engineering domains. In specific, 21 (35%) subjects indicated their current organization to be affiliated with multiple domains and the rest expressed to be associated with single domain. Table 2 presents the frequency of respondents associated with various domains. Among the 21 subjects associated with multiple domains, we observed 11 subjects to be linked with more than two domains and four subjects were associated with at least five domains. From Table 2, we can further notice that most of the respondents (51.6%) indicated to be associated with ICT/ telecommunication domain (DC13), followed by, five respondents (8.3%) associated with Web applications (DC17) domain.

**Table 2 Frequency of respondents associated with various software engineering domains**

| Domains' combination ID | Domain | | | | | | | | | | | | | Frequency |
|---|---|---|---|---|---|---|---|---|---|---|---|---|---|---|
| | Finance/ Banking/ Insurance | (ICT)/ Telecom | Infrastructure services | Mobile applications | Web applications | E-learning | Embedded systems | Logistics/ Shipping | Automotive | Entertainment/ Recreation | Government/ Military | Internet | E-commerce | |
| DC1 | ✓ | ✓ | ✓ | ✓ | ✓ | | ✓ | | ✓ | | | | ✓ | 1 |
| DC2 | ✓ | ✓ | | | ✓ | | ✓ | | | | | ✓ | | 1 |
| DC3 | ✓ | | ✓ | | | | | | | | | ✓ | | 1 |
| DC4 | | ✓ | ✓ | ✓ | ✓ | | | | | | | | | 1 |
| DC5 | | ✓ | ✓ | | | | ✓ | | | | | ✓ | | 1 |
| DC6 | | ✓ | ✓ | | | | | | | | | | | 2 |

| Domains' combination ID | Domain | | | | | | | | | | | | | Frequency |
|---|---|---|---|---|---|---|---|---|---|---|---|---|---|---|
| | Finance/ Banking/ Insurance | (ICT)/ Telecom | Infrastructure services | Mobile applications | Web applications | E-learning | Embedded systems | Logistics/ Shipping | Automotive | Entertainment/ Recreation | Government/ Military | Internet | E-commerce | |
| DC7 | | ✓ | | ✓ | ✓ | | | | | | | | | 2 |
| DC8 | | ✓ | | ✓ | | | ✓ | | | | | ✓ | ✓ | 1 |
| DC9 | | ✓ | | | ✓ | | | | | ✓ | | ✓ | | 1 |
| DC10 | | ✓ | | | | ✓ | | | | | | | | 1 |
| DC11 | | ✓ | | | | | ✓ | | ✓ | | | | | 1 |
| DC12 | | ✓ | | | | | | ✓ | ✓ | ✓ | ✓ | ✓ | ✓ | 1 |
| DC13 | | ✓ | | | | | | | | | | | | 31 |
| DC14 | | | ✓ | | | | | | | | | ✓ | | 1 |
| DC15 | | | | ✓ | ✓ | | | | | | | | | 1 |
| DC16 | | | | ✓ | | | | | | | | | | 2 |
| DC17 | | | | | ✓ | | | | | | | | | 5 |
| DC18 | | | | | | | ✓ | | | | | | | 2 |
| DC19 | | | | | | | | ✓ | | | | | | 2 |
| DC20 | | | | | | | | | ✓ | | | | | 1 |
| DC21 | | | | | | | | | | | | | ✓ | 1 |

Further, by examining the ASD methodologies employed in the subjects' work environment, we observed that 85% of the subjects adhered to one methodology (see Table 3). In the cases where subjects exclusively adhered to a single methodology, we noticed that the majority (68.3%) practiced Scrum (MC4), followed by Kanban (MC7: 13.3%), Adaptive Software Development (AdSD, MC5: 1.6%) and Feature Driven Development (FDD, MC6: 1.6%). Interestingly, Scrum was also seen to be practiced among all the nine subjects who reported practicing multiple methodologies (MC1, MC2 and MC3). Scrum in combination with Kanban (MC3: Scrumban) was reported to be practiced by 8.3% subjects and a combination of Scrum and FDD (MC2) were reported to be practiced by 5% subjects.

**Table 3. Software development methodologies practiced at subjects' work environment**

| | ASD methodology | | | | | Frequency |
|---|---|---|---|---|---|---|
| **Methods' combination ID** | **Scrum** | **DSDM** | **AdSD** | **FDD** | **Kanban** | |
| MC1 | ✓ | ✓ | | ✓ | | 1 |
| MC2 | ✓ | | | ✓ | | 3 |
| MC3 | ✓ | | | | ✓ | 5 |
| MC4 | ✓ | | | | | 41 |
| MC5 | | | ✓ | | | 1 |
| MC6 | | | | ✓ | | 1 |
| MC7 | | | | | ✓ | 8 |

The information on number of members working in a team, alongside the subjects, has been displayed in Figure 2.1. It shows that more than 80% of the subjects reported to be working in teams whose size was less than 10. While 55% of the subjects worked in teams having six to 10 members, a little more than a quarter of the subjects (26.7%) worked in teams having less than or equal to five members. The reason for these team sizes could be attributed to the ASD methodologies practiced at subjects' workplace. From Table 3, we can observe that a large proportion of the subjects practiced Scrum in their teams and in general, a standard Scrum team consists of five to nine members [80], [81]. On the other hand, around 18.3% of the subjects worked in teams with more than 10 members and a major portion of these subjects were part of 11 to 15 member teams. Further, two subjects (3.3%) reported to be a part of large team having 21 to 25 members. It is worth noting that all the subjects indicated the number of members working in their teams. Thus, with the experience of working in teams, we believe that each subject would have developed certain level of consciousness and knowledge that enables them to reflect upon the relevance of team capability measures.

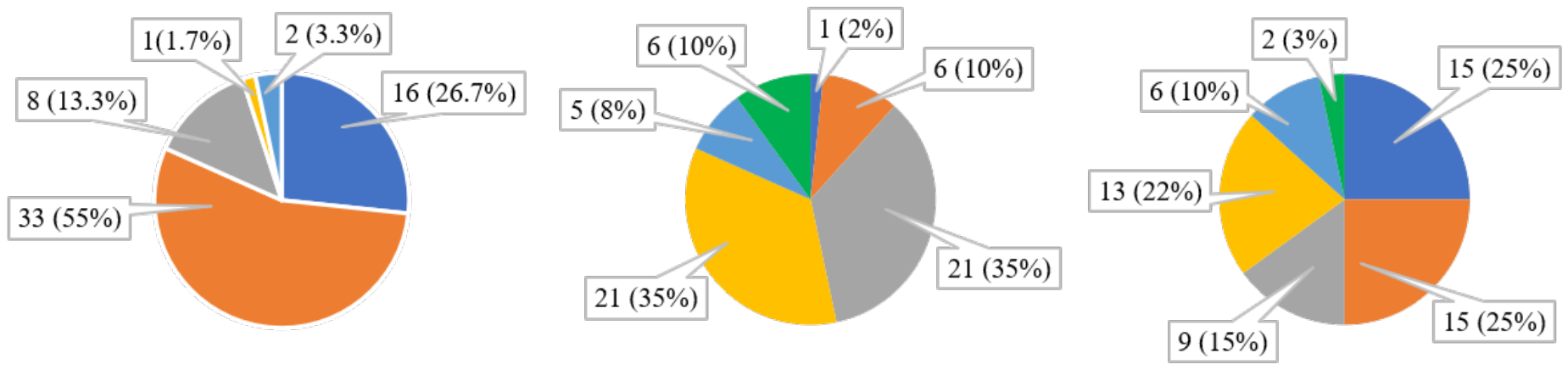

**Figure 2.1. Distribution of members (m) in subjects' teams**

**Figure 2.2. Total work experience (w) of subjects in years (y)**

**Figure 2.3. Years (y) of experience (e) in current organization**

| **Figure** | | | | | | |
|---|---|---|---|---|---|---|
| Figure 2.1 | [1, 5] m | [6, 10] m | [11, 15] m | [16, 20] m | [21, 25] m | |
| Figure 2.2 | w < 3 y | 3 ≤ w < 4 y | 4 ≤ w < 5 y | 5 ≤ w < 6 y | 6 ≤ w < 7 y | w ≥ 7 y |
| Figure 2.3 | e < 2 y | 2 ≤ e < 3 y | 3 ≤ e < 4 y | 4 ≤ e < 5 y | 5 ≤ e < 6 y | e ≥ 6 y |

**Figure 2. Pie charts illustrating the distributions across various categories**

Looking at the distribution of subjects' work experience, as shown in Figure 2.2, we noticed that our sample consisted of a mix of professionals with different levels of experience. While only one of the subjects had less than three years of experience, the total work experience of 10% of the subjects was under the range of three to less than four years. The proportion of respondents whose work experience was under the range of four to less than five years was the same (35%) as the case where the respondents had five to less than six years of experience. Around 8% of the subjects had six to less than seven years of experience. On the other hand, 10% of the subjects had more than seven years of experience. Among this 10%, two subjects had the highest work experience ranging from 13 to 15 years. The median value of overall industrial experience, for all the recruited subjects in our study, was five years and we can ascertain that, on the whole, it was an experienced sample consisting of midlevel to senior level professionals [82].

All the subjects also stated the number of years spent in their current company, indicating that they were currently employed and that they were currently working in industry. The distribution of responses with respect to experience in the current organization are presented in Figure 2.3. From the figure, we can clearly observe that 50% of the subjects had worked at their current organization for less than three years. Whereas the other half of the respondents were associated with their current organization for three or more years. It is important to note that all the subjects had an experience of working in their current organization for at least one year.

Among the 60 respondents, 41 members (68.3%) undertook the primary role of a developer, followed by groups of five members for the roles of continuous integration engineer (8.3%), Scrum master (8.3%) and tester (8.3%). Among the aforementioned top four roles, two roles (developer and Scrum master) were noticed to be Scrum team specific [80]. Besides these, single responses were recorded for the roles of team leader, technical expert, architect and DevOps engineer. All the roles were observed to pertain to software development, and this affirms that the subjects were currently working in a software organization.

# 4. Results

This section presents the results of our statistical analysis. The relevance levels for various capability measures indicated by survey respondents were analyzed to answer the research questions.

## 4.1. Inspecting the perceived relevance of capability measures (RQ.1)

To investigate what measures practitioners perceive as relevant for representing capability in ASD, we examined how the individual and team capability measures were rated by the survey respondents. We started by analyzing the responses for the set of capability measures that fall under the professional category, followed by social and innovative categories. Under each of the three categories, the multi-dimensional Likert data associated with capability measures was tabulated and displayed in the form of a heatmap [83]. Heatmap facilitates visualizing the Likert data and highlights the predominant cells in a table. This helps in identifying which measures were perceived as highly relevant and irrelevant; and also aids in segregating the rows of capability measures with similar response frequency. In order to generate a heatmap, we used a color gradient function that sets the lowest possible value (0%) to white, the highest possible value (100%) to dark blue, and mid-range values to a corresponding transition between the extremes.

For the sake of comparing the responses of various capability measures and facilitating easy lookup of any measure's description from the supplementary material, each capability measure was assigned a unique identifier such as IC.X.Y.Z or TC.X.Y.Z, where IC stands for Individual Capability measure and TC stands for Team Capability measure. X represents the primary category under which a capability measure is classified, and it corresponds to P for Professional, S for Social and I for innovative categories. Next, Y represents the sub-category under which a capability measure is classified. In the light of the multiple sub-categories that fall under a primary category, the sub-categories are assigned a serial number and Y corresponds to the serial number of a sub-category. Finally, Z is a reference number for each capability measure that is classified under a sub-category Y.

Each heatmap displays the list of capability measures in the order of their appearance in our questionnaire. Further, for each capability measure, the percentage of responses received over the Likert item (Highly Relevant (HR), Relevant (R), Somewhat Relevant (SR), Irrelevant (IR) and I Don't Know (IDK)) is presented. The figures of percentage of response were further used towards the collective analysis of capability measures across the three primary categories. This was accomplished via a Filtering and Sorting and Criteria (FSC), the details of which will be discussed in the next sub-section, while presenting the first heatmap.

### 4.1.1. Perceived relevance of individual capability measures (RQ.1.1)

We present three heatmaps for displaying the responses in relation to each of the three primary categories of individual capability measures. In order to figure out which measures were perceived as HR/R within each primary category, we first filtered the measures which were familiar to at least 95% of the respondents ($p(IDK) \leq 5\%$) and then subsequently filtered the ones recognized by majority of the members as relevant ($p(HR + R) > p(SR + IR)$). The resultant measures were then ranked by sorting in terms of HR response frequency (descending), followed by R (descending), SR (ascending), IR (ascending) and IDK (ascending) response frequencies. We will refer to these criteria as FSC-1. The ranks of the resultant measures within each primary category are presented in the respective heatmap table under the column FSC-1. Please note that the measures which were recognized by the majority respondents as trivial are marked as '*' under the column FSC-1.

Based on the ranks and response frequencies within each primary category, we examined which capability measures were rated as most (and least) relevant. Besides this, we also inspected the perceptions of agile practitioners with respect to the ASD-specific capability measures that were widely discussed in the SE literature. This was facilitated by segregating measures from our SLR [22] that were emphasized by multiple primary studies. The details of these analyses are presented next.

#### 4.1.1.1. Insights from the professional category

The heatmap from Table 4 displays the perceived relevance of individual capability measures under the professional category.

- The top five measures which were indicated as HR by at least 43% of the respondents:
  - Software construction (*debugging and testing* [IC.P.4.7], *integrating and collaborating* [IC.P.4.8], and *system integration and verification* [IC.P.4.1])
  - Software security and safety (*testing* [IC.P.10.5] and *quality* [IC.P.10.3])
- The measures which were perceived by most respondents as not relevant ($p(HR + R) < p(SR + IR)$):
  - Around 58% of the respondents considered the number of years spent by a person in an organization (*years in company* [IC.P.13.2]) to be trivial ((p(SR)+p(IR)) for representing the capability of an agile team member.
- The list of measures that were not familiar to more than 5% of the respondents:
  - All the measures associated with software sustainment sub-category
  - All the measures associated with software measurement sub-category
  - Software process model and life cycle model (*software development life cycle models* [IC.P.5.1], *process definition and tailoring* [IC.P.5.2] and *process assessment and improvement* [IC.P.5.4])

- Details of how agile practitioners perceived the relevance of widely discussed measures in the SE literature [22]:
  - The measure concerning experience of working in different roles over past years (*prior work experience* [IC.P.13.1]) was discussed by three studies [39], [84], [85] and it was considered as relevant (p(HR)+p(R)) by around 58% of the respondents, among whom, 20% regarded it as HR.
  - In relation to software process model and life cycle model, the *process implementation and management* [IC.P.5.3] proficiency was discussed by two studies [84], [85]. This measure was considered as relevant by around 71% of the respondents (HR=23%).
  - *Programming experience* [IC.P.13.3] of a person which was discussed by two studies [84], [85], was considered as relevant by around 76% of the respondents (HR=30%).
  - The ability to organize activities for achieving goals (*planning skills* [IC.P.13.5]) was discussed by two studies [40], [85] from our SLR. It was considered as relevant by around 75% of the respondents (HR=38%).

**Table 4. Heatmap of response frequencies for individual capability measures; Primary category: Professional. The IDs of measures with FSC-2 ranks are highlighted in bold**

| Sub-category | ID | Capability measure | Percentage (n=60) | | | | | FSC-1 | FSC-2 |
|---|---|---|---|---|---|---|---|---|---|
| | | | *HR* | *R* | *SR* | *IR* | *IDK* | *Rank* | *Rank* |
| Software requirements | IC.P.1.1 | Elicitation | 30 | 46.67 | 20 | 1.67 | 1.67 | #22 | |
| | IC.P.1.2 | Analysis | 38.33 | 48.33 | 10 | 3.33 | 0 | #9 | |
| | IC.P.1.3 | Specification | 35 | 53.33 | 8.33 | 3.33 | 0 | #14 | |
| | IC.P.1.4 | Verification | 30 | 50 | 16.67 | 3.33 | 0 | #19 | |
| Software design | IC.P.2.1 | Fundamentals | 30 | 48.33 | 16.67 | 3.33 | 1.67 | #20 | |
| | IC.P.2.2 | Strategies and methods | 25 | 51.67 | 21.67 | 1.67 | 0 | #30 | |
| | IC.P.2.3 | Software architectural design | 28.33 | 50 | 15 | 5 | 1.67 | #23 | |
| | IC.P.2.4 | Quality analysis and evaluation | 31.67 | 43.33 | 16.67 | 8.33 | 0 | #16 | |
| Software system engineering | IC.P.3.1 | Concept definition | 28.33 | 38.33 | 26.67 | 6.67 | 0 | #27 | |
| | IC.P.3.2 | System development life cycle modeling | 23.33 | 40 | 26.67 | 8.33 | 1.67 | #37 | |
| | IC.P.3.3 | Software-intensive systems engineering | 15 | 35 | 36.67 | 8.33 | 5 | #44 | |
| | IC.P.3.4 | System design | 25 | 48.33 | 20 | 6.67 | 0 | #31 | |
| | IC.P.3.5 | Requirements allocation and flow-down | 26.67 | 43.33 | 20 | 6.67 | 3.33 | #29 | |
| | IC.P.3.6 | Component engineering | 23.33 | 40 | 23.33 | 6.67 | 6.67 | | |
| Software construction | IC.P.4.1 | System integration and verification | 43.33 | 38.33 | 13.33 | 3.33 | 1.67 | #5 | |
| | IC.P.4.2 | System validation and deployment | 41.67 | 38.33 | 16.67 | 3.33 | 0 | #7 | |
| | IC.P.4.3 | System sustainment planning | 18.33 | 35 | 41.67 | 3.33 | 1.67 | #42 | |
| | IC.P.4.4 | Software construction planning | 23.33 | 48.33 | 18.33 | 6.67 | 3.33 | #34 | |
| | IC.P.4.5 | Managing software construction | 21.67 | 45 | 18.33 | 13.33 | 1.67 | #38 | |
| | **IC.P.4.6** | Detailed design and coding | 43.33 | 38.33 | 16.67 | 1.67 | 0 | #6 | #22 |
| | **IC.P.4.7** | Debugging and testing | 48.33 | 41.67 | 6.67 | 3.33 | 0 | #1 | #18 |
| | IC.P.4.8 | Integrating and collaborating | 46.67 | 46.67 | 5 | 0 | 1.67 | #2 | |
| Software process model and life cycle model | IC.P.5.1 | Software development life cycle models | 23.33 | 43.33 | 21.67 | 3.33 | 8.33 | | |
| | IC.P.5.2 | Process definition and tailoring | 13.33 | 45 | 25 | 8.33 | 8.33 | | |
| | IC.P.5.3 | Process implementation and management | 23.33 | 48.33 | 20 | 3.33 | 5 | #36 | |
| | IC.P.5.4 | Process assessment and improvement | 21.67 | 40 | 28.33 | 3.33 | 6.67 | | |
| Human-computer interaction | IC.P.6.1 | Requirements | 28.33 | 46.67 | 8.33 | 13.33 | 3.33 | #24 | |
| | IC.P.6.2 | Interaction style design | 11.67 | 50 | 23.33 | 10 | 5 | #45 | |
| | IC.P.6.3 | Visual design | 11.67 | 40 | 28.33 | 11.67 | 8.33 | | |
| | IC.P.6.4 | Usability testing | 28.33 | 43.33 | 11.67 | 13.33 | 3.33 | #25 | |
| | IC.P.6.5 | Accessibility | 28.33 | 43.33 | 11.67 | 8.33 | 8.33 | | |
| Software testing | IC.P.7.1 | Techniques | 36.67 | 46.67 | 10 | 5 | 1.67 | #12 | |
| | IC.P.7.2 | Planning | 26.67 | 48.33 | 18.33 | 6.67 | 0 | #28 | |
| | IC.P.7.3 | Infrastructure | 25 | 43.33 | 26.67 | 3.33 | 1.67 | #33 | |
| | IC.P.7.4 | Measurement & defect tracking | 40 | 36.67 | 13.33 | 6.67 | 3.33 | #8 | |
| Software quality | IC.P.8.1 | Independent process and product audits | 11.67 | 45 | 25 | 8.33 | 10 | | |
| | IC.P.8.2 | Statistical control | 20 | 38.33 | 25 | 11.67 | 5 | #40 | |

| Sub-category | ID | Capability measure | Percentage (n=60) | | | | | FSC-1 | FSC-2 |
|---|---|---|---|---|---|---|---|---|---|
| | | | *HR* | *R* | *SR* | *IR* | *IDK* | *Rank* | *Rank* |
| | IC.P.8.3 | Management | 15 | 46.67 | 23.33 | 11.67 | 3.33 | #43 | |
| | IC.P.8.4 | Reviews, walkthroughs and inspections | 38.33 | 38.33 | 16.67 | 6.67 | 0 | #10 | |
| Software sustainment | IC.P.9.1 | Software transition | 13.33 | 43.33 | 30 | 3.33 | 10 | | |
| | IC.P.9.2 | Software support | 18.33 | 45 | 23.33 | 5 | 8.33 | | |
| | IC.P.9.3 | Software maintenance | 23.33 | 43.33 | 18.33 | 6.67 | 8.33 | | |
| Software security and safety | IC.P.10.1 | Design | 36.67 | 41.67 | 16.67 | 3.33 | 1.67 | #13 | |
| | IC.P.10.2 | Construction | 23.33 | 48.33 | 18.33 | 6.67 | 3.33 | #35 | |
| | **IC.P.10.3** | Quality | 45 | 40 | 11.67 | 3.33 | 0 | #4 | #21 |
| | IC.P.10.4 | Requirements | 33.33 | 43.33 | 15 | 8.33 | 0 | #15 | |
| | IC.P.10.5 | Testing | 46.67 | 30 | 15 | 6.67 | 1.67 | #3 | |
| | IC.P.10.6 | Process | 31.67 | 36.67 | 18.33 | 10 | 3.33 | #18 | |
| Software configuration management | IC.P.11.1 | Plan software configuration management | 31.67 | 43.33 | 18.33 | 1.67 | 5 | #17 | |
| | IC.P.11.2 | Conduct software configuration management | 25 | 45 | 21.67 | 3.33 | 5 | #32 | |
| | IC.P.11.3 | Manage software releases | 28.33 | 41.67 | 23.33 | 3.33 | 3.33 | #26 | |
| Software measurement | IC.P.12.1 | Plan software measurement process | 25 | 38.33 | 25 | 3.33 | 8.33 | | |
| | IC.P.12.2 | Perform software measurement process | 21.67 | 41.67 | 23.33 | 1.67 | 11.67 | | |
| Miscellaneous | IC.P.13.1 | Prior work experience | 20 | 38.33 | 25 | 15 | 1.67 | #41 | |
| | IC.P.13.2 | Years in company | 11.67 | 30 | 33.33 | 25 | 0 | * | |
| | IC.P.13.3 | Programming experience | 30 | 46.67 | 16.67 | 5 | 1.67 | #21 | |
| | IC.P.13.4 | Allocated full-time | 21.67 | 36.67 | 25 | 16.67 | 0 | #39 | |
| | IC.P.13.5 | Planning skills | 38.33 | 36.67 | 18.33 | 3.33 | 3.33 | #11 | |

#### 4.1.1.2. Insights from the social category

The heatmap from Table 5 displays the perceived relevance of individual capability measures under the social category.

- The top five measures which were indicated as HR by at least 63% of respondents:
  - Work ethics (*responsibility* [IC.S.5.5] and *motivation to work* [IC.S.5.3])
  - Communication (*listening skills* [IC.S.2.2] and *questioning skills* [IC.S.2.3])
  - Affective (*team participation skills* [IC.S.1.9])
- The measures which were perceived by most respondents as trivial:
  - Without much surprise, the characteristic of directing one's interest inwards towards one's own thoughts and not being social (*introversion* [IC.S.4.2]) was considered by 68% of the respondents as not relevant in relation to representing an agile team member's capability.
  - Around 50% of the respondents considered a person's quality of deciding based on social considerations (*feeling* [IC.S.4.7]) to be trivial.
- The list of measures that were not familiar to more than 5% of the respondents:
  - Interpersonal (*willingness to confront* [IC.S.3.6])
  - Personal (*extroversion* [IC.S.4.3] and *tenacity* [IC.S.4.14])
- Details of how agile practitioners perceived the relevance of widely discussed measures in the SE literature [22]:
  - In relation to communication skills of a person, proficiency in *oral communication* [IC.S.2.1], *listening skills* [IC.S.2.2] and *questioning skills* [IC.S.2.3] were discussed by four studies [39], [85], [86], [87]. The dominant measure among them was *listening skills*. It was considered as relevant by around 95% of the respondents, among whom, 71% regarded it as HR. Next to *listening skills*, we observed that the *questioning skills* measure was considered as relevant by all the respondents, where 63% of them considered it as HR. Next, *oral communication* was considered as relevant by around 88% of the respondents (HR=53%).
  - In relation to interpersonal skills of a person, the measure concerning *attitude* of professionals [IC.S.3.4] was discussed by four studies [39], [84], [85], [87], and this measure was considered as relevant by around 90% of the respondents (HR=58%). Further, a person's state of being prepared to work collaboratively with a group, towards achieving a common goal (*team work oriented* [IC.S.3.5]) was discussed by three studies [86], [87], [85]. This measure was considered as relevant by around 95% of the respondents (HR=63%).

- A person's quality of being outgoing (*extroversion* [IC.S.4.3]) was discussed by three studies [20], [21], [41]. We have noticed mixed opinions for this measure. The proportion of respondents that considered *extroversion* to be relevant was same as the proportion that considered it trivial. Around 11% of the respondents indicated *extroversion* to be HR for characterizing the capability of an agile team member.

**Table 5. Heatmap of response frequencies for individual capability measures; Primary category: Social. The IDs of measures with FSC-2 ranks are highlighted in bold**

| Sub-category | ID | Capability measure | Percentage | | | | | FSC-1 | FSC-2 |
|---|---|---|---|---|---|---|---|---|---|
| | | | *HR* | *R* | *SR* | *IR* | *IDK* | *Rank* | *Rank* |
| Affective | IC.S.1.1 | Aptitude | 28.33 | 45 | 20 | 5 | 1.67 | #36 | |
| | IC.S.1.2 | Initiative | 45 | 46.67 | 6.67 | 1.67 | 0 | #22 | |
| | **IC.S.1.3** | Enthusiasm | 46.67 | 41.67 | 10 | 1.67 | 0 | #21 | #20 |
| | **IC.S.1.4** | Work ethic | 51.67 | 40 | 8.33 | 0 | 0 | #16 | #13 |
| | **IC.S.1.5** | Willingness | 55 | 38.33 | 5 | 1.67 | 0 | #11 | #10 |
| | IC.S.1.6 | Planning skills | 38.33 | 38.33 | 20 | 1.67 | 1.67 | #28 | |
| | IC.S.1.7 | Trustworthiness | 40 | 53.33 | 5 | 0 | 1.67 | #24 | |
| | IC.S.1.8 | Non-technical leadership skills | 11.67 | 43.33 | 35 | 5 | 5 | #46 | |
| | **IC.S.1.9** | Team participation skills | 63.33 | 35 | 1.67 | 0 | 0 | #5 | #4 |
| | IC.S.1.10 | Technical leadership skills | 38.33 | 40 | 16.67 | 5 | 0 | #27 | |
| Communication | IC.S.2.1 | Oral communication | 53.33 | 35 | 10 | 0 | 1.67 | #15 | |
| | **IC.S.2.2** | Listening skills | 71.67 | 23.33 | 5 | 0 | 0 | #2 | #2 |
| | **IC.S.2.3** | Questioning skills | 63.33 | 36.67 | 0 | 0 | 0 | #4 | #3 |
| Interpersonal | **IC.S.3.1** | Seeks help | 48.33 | 41.67 | 8.33 | 1.67 | 0 | #18 | #17 |
| | **IC.S.3.2** | Helps others | 48.33 | 43.33 | 6.67 | 1.67 | 0 | #17 | #16 |
| | IC.S.3.3 | Customer orientation | 23.33 | 43.33 | 30 | 3.33 | 0 | #41 | |
| | IC.S.3.4 | Attitude | 58.33 | 31.67 | 6.67 | 1.67 | 1.67 | #9 | |
| | **IC.S.3.5** | Teamwork oriented | 63.33 | 31.67 | 5 | 0 | 0 | #6 | #5 |
| | IC.S.3.6 | Willingness to confront | 40 | 40 | 10 | 3.33 | 6.67 | | |
| Personal | IC.S.4.1 | Driven by desire to contribute | 55 | 35 | 8.33 | 0 | 1.67 | #12 | |
| | IC.S.4.2 | Introversion | 1.67 | 25 | 33.33 | 35 | 5 | * | |
| | IC.S.4.3 | Extroversion | 11.67 | 35 | 26.67 | 20 | 6.67 | | |
| | IC.S.4.4 | Sensing | 18.33 | 45 | 26.67 | 5 | 5 | #44 | |
| | IC.S.4.5 | Intuitive | 21.67 | 53.33 | 20 | 1.67 | 3.33 | #42 | |
| | IC.S.4.6 | Thinking | 40 | 48.33 | 10 | 1.67 | 0 | #25 | |
| | IC.S.4.7 | Feeling | 13.33 | 35 | 38.33 | 11.67 | 1.67 | * | |
| | IC.S.4.8 | Judging | 16.67 | 46.67 | 16.67 | 18.33 | 1.67 | #45 | |
| | IC.S.4.9 | Perceiving | 20 | 50 | 26.67 | 3.33 | 0 | #43 | |
| | IC.S.4.10 | Openness | 36.67 | 51.67 | 8.33 | 3.33 | 0 | #32 | |
| | IC.S.4.11 | Conscientiousness | 28.33 | 53.33 | 15 | 0 | 3.33 | #35 | |
| | IC.S.4.12 | Agreeableness | 25 | 53.33 | 13.33 | 6.67 | 1.67 | #39 | |
| | IC.S.4.13 | Charisma | 8.33 | 41.67 | 30 | 16.67 | 3.33 | #47 | |
| | IC.S.4.14 | Tenacity | 15 | 50 | 21.67 | 5 | 8.33 | | |
| | IC.S.4.15 | Behavior | 43.33 | 35 | 16.67 | 3.33 | 1.67 | #23 | |
| | **IC.S.4.16** | Knowledge | 53.33 | 36.67 | 8.33 | 1.67 | 0 | #13 | #11 |
| | IC.S.4.17 | Education | 38.33 | 35 | 18.33 | 8.33 | 0 | #29 | |
| | IC.S.4.18 | Pride in quality and productivity | 40 | 33.33 | 20 | 6.67 | 0 | #26 | |
| | IC.S.4.19 | Perseverance | 38.33 | 35 | 21.67 | 0 | 5 | #30 | |
| | **IC.S.4.20** | Desire to improve things | 58.33 | 33.33 | 6.67 | 1.67 | 0 | #8 | #9 |
| | IC.S.4.21 | Pro-active/ initiator/ driver | 46.67 | 46.67 | 6.67 | 0 | 0 | #19 | |
| | IC.S.4.22 | Maintaining "big picture" view | 46.67 | 43.33 | 6.67 | 1.67 | 1.67 | #20 | |
| | IC.S.4.23 | Desire to do/bias for action | 33.33 | 30 | 28.33 | 3.33 | 5 | #34 | |
| | IC.S.4.24 | Thoroughness | 35 | 43.33 | 15 | 3.33 | 3.33 | #33 | |
| | IC.S.4.25 | Sense of mission | 25 | 60 | 11.67 | 1.67 | 1.67 | #38 | |
| | IC.S.4.26 | Strength of convictions | 25 | 45 | 23.33 | 5 | 1.67 | #40 | |
| | IC.S.4.27 | Mixes personal and work goals | 21.67 | 26.67 | 15 | 31.67 | 5 | #48 | |
| | IC.S.4.28 | Pro-active role with management | 28.33 | 43.33 | 20 | 5 | 3.33 | #37 | |
| Work ethics | IC.S.5.1 | Flexibility | 36.67 | 51.67 | 6.67 | 3.33 | 1.67 | #31 | |
| | **IC.S.5.2** | Time management | 53.33 | 36.67 | 8.33 | 1.67 | 0 | #14 | #12 |
| | IC.S.5.3 | Motivation to work | 68.33 | 23.33 | 6.67 | 0 | 1.67 | #3 | |
| | **IC.S.5.4** | Commitment | 60 | 38.33 | 0 | 1.67 | 0 | #7 | #6 |
| | **IC.S.5.5** | Responsibility | 75 | 25 | 0 | 0 | 0 | #1 | #1 |
| | IC.S.5.6 | Integrity/ honesty/ ethics | 56.67 | 36.67 | 3.33 | 1.67 | 1.67 | #10 | |

#### 4.1.1.3. Insights from the innovative category

The heatmap from Table 6 displays the perceived relevance of individual capability measures under the innovative category.

- The top five measures which were indicated as HR by at least 50% of the respondents:
  - Enterprising (*identifying problem* [IC.I.2.1], *seeking improvement* [IC.I.2.2], and *gathering and evaluating information* [IC.I.2.3])
  - Integrating perspectives (*openness to ideas* [IC.I.3.1] and *collaborating* [IC.I.3.3])
- The list of measures that were not familiar to more than 5% of the respondents:
  - Forecasting (*sensitivity to situations* [IC.I.4.4])
  - Managing change (*challenging the status quo* [IC.I.5.1] and *reinforcing change* [IC.I.5.3])
- Details of how agile practitioners perceived the relevance of widely discussed measures in SE literature [22]:
  - In relation to creativity of a person, the measures concerning the ability to think carefully about an idea (*critical thinking* [IC.I.1.2]), and the ability to find a solution to a problem using imagination (*creative problem solving* [IC.I.1.4]) were discussed by two studies [86], [87]. The dominant measure among them was *creative problem solving*. It was considered as relevant by around 86% of the respondents, among whom, 48% considered it to be HR. Whereas *critical thinking* was considered as relevant by around 90% of the respondents (HR= 41%).

**Table 6. Heatmap of response frequencies for individual capability measures; Primary category: Innovative. The IDs of measures with FSC-2 ranks are highlighted in bold**

| Sub-category | ID | Capability measure | Percentage | | | | | FSC-1 | FSC-2 |
|---|---|---|---|---|---|---|---|---|---|
| | | | ***HR*** | ***R*** | ***SR*** | ***IR*** | ***IDK*** | ***Rank*** | ***Rank*** |
| Creativity | IC.I.1.1 | Generating ideas | 38.33 | 48.33 | 10 | 1.67 | 1.67 | #11 | |
| | IC.I.1.2 | Critical thinking | 41.67 | 48.33 | 8.33 | 1.67 | 0 | #8 | |
| | IC.I.1.3 | Synthesis/reorganization | 16.67 | 51.67 | 21.67 | 8.33 | 1.67 | #18 | |
| | **IC.I.1.4** | Creative problem solving | 48.33 | 38.33 | 10 | 3.33 | 0 | #6 | #19 |
| | IC.I.1.5 | Attention to detail | 43.33 | 36.67 | 18.33 | 0 | 1.67 | #7 | |
| Enterprising | **IC.I.2.1** | Identifying problem | 58.33 | 36.67 | 3.33 | 1.67 | 0 | #2 | #8 |
| | IC.I.2.2 | Seeking improvement | 51.67 | 36.67 | 10 | 0 | 1.67 | #3 | |
| | **IC.I.2.3** | Gathering and evaluating information | 51.67 | 35 | 13.33 | 0 | 0 | #4 | #14 |
| | IC.I.2.4 | Independent thinking | 40 | 43.33 | 16.67 | 0 | 0 | #10 | |
| | IC.I.2.5 | Technological savvy | 25 | 40 | 33.33 | 0 | 1.67 | #16 | |
| Integrating perspectives | **IC.I.3.1** | Openness to ideas | 60 | 31.67 | 8.33 | 0 | 0 | #1 | #7 |
| | IC.I.3.2 | Research orientation | 41.67 | 41.67 | 13.33 | 3.33 | 0 | #9 | |
| | **IC.I.3.3** | Collaborating | 50 | 45 | 5 | 0 | 0 | #5 | #15 |
| | IC.I.3.4 | Engaging in non-work related interests | 18.33 | 38.33 | 26.67 | 13.33 | 3.33 | #17 | |
| Forecasting | IC.I.4.1 | Evaluating long-term consequences | 36.67 | 38.33 | 21.67 | 0 | 3.33 | #12 | |
| | IC.I.4.2 | Visioning | 31.67 | 48.33 | 13.33 | 1.67 | 5 | #13 | |
| | IC.I.4.3 | Managing the future | 26.67 | 41.67 | 23.33 | 3.33 | 5 | #15 | |
| | IC.I.4.4 | Sensitivity to Situations | 25 | 35 | 20 | 13.33 | 6.67 | | |
| Managing change | IC.I.5.1 | Challenging the status quo | 21.67 | 51.67 | 18.33 | 1.67 | 6.67 | | |
| | IC.I.5.2 | Intelligent risk-taking | 28.33 | 48.33 | 21.67 | 0 | 1.67 | #14 | |
| | IC.I.5.3 | Reinforcing change | 21.67 | 46.67 | 20 | 3.33 | 8.33 | | |

#### 4.1.1.4. Key insights from the three primary categories

A bird's-eye view of the heatmaps, clearly indicate that a major portion of the measures were perceived as relevant for characterizing the capability of a software professional. This becomes even more evident when the capability measures (from Table 4, Table 5 and Table 6) are collated, and the percentages (p) presented in the HR and R columns are aggregated (percentage of subjects perceiving a measure as relevant). We saw that 127 out of 132 individual capability measures were indicated as HR or R (HR/R) by more than 50% of the respondents ($p(HR + R) > p(SR + IR)$).

All the 21 measures classified under innovative category were indicated as HR/R by at least 60% of the respondents (see Table 6). Whereas in the case of professional category, all except for *years in company* [IC.P.13.2], i.e. 57 measures, were indicated as HR/R by more than half of the respondents (see Table 4) and under social category, all except for *introversion* [IC.S.4.2], *extroversion* [IC.S.4.3], *feeling* [IC.S.4.7], and *mixes personal and work goals* [IC.S.4.27], i.e. 49 measures were also perceived to be HR/R by more than 50% of the respondents (see Table 5).

Upon a closer inspection, we observed that two capability measures were not only unanimously indicated to be relevant (p(SR), p(IR) and p(IDK) = 0%), but also were perceived to be HR by most respondents. These measures were *responsibility* [IC.S.5.5] and *questioning skills* [IC.S.2.3].

For the sake of collectively analyzing measures from the three primary categories, we employed a FSC which first filter measures that were widely known (p(IDK) = 0%) as well as perceived as HR by the majority of the respondents. Next, the filtered measures were ranked by sorting in terms of HR response frequency (descending) followed by IR response frequency (ascending). We will refer to these criteria as FSC-2 hence forth and it resulted in identifying 22 measures across the three primary categories. The ranks of these measures are presented in Table 4, Table 5 and Table 6 under the column FSC-2 and the IDs of these measures are highlighted in bold. These 22 measures were perceived as highly relevant for characterizing the capability of an individual, by at least 43% of the respondents. We can notice that almost two-thirds of these 22 measures correspond to the social category, followed by five measures from the innovative category and three measures from the professional category.

In fact, the top five ranked capability measures among the FSC-2 results belonged to social category, where at least 63% of the respondents consider them (*responsibility*, *listening skills*, *questioning skills*, *team participation skills* and *being teamwork oriented*) as very crucial. This clearly indicates that non-technical abilities associated with emotions, work ethics, communication and interpersonal skills, were highly regarded by agile practitioners as capabilities that are relevant for agile team members. Next to social category, the measures from innovative category pertaining to enterprising (*identifying problem* and *gathering and evaluating information*), integrating perspectives (*openness to ideas* and *collaborating*) and creativity (*creative problem solving*) were also highly regarded. Within the professional category, measures associated with software construction (*detailed design and coding* and *debugging and testing*) and *software security and safety quality* were opined to be crucial.

In order to present a summary of how agile practitioners perceived the relevance of individual capability measures, especially with regard to the ones that were widely discussed in the SE literature, a Venn diagram is presented in Figure 3. Besides portraying measures from each primary category, the Venn diagram also presents the measures that were overlapping between the primary categories. Former studies [22] discussed *planning skills* within professional and social contexts ([IC.P.13.5] and [IC.S.1.6]). Whereas *teamwork oriented* [IC.S.3.5] and *collaborating* [IC.I.3.3] were discussed in social and innovative contexts, respectively. However, in both cases, we observed that the response frequencies for overlapping measures were almost the same. From Figure 3, we can clearly see that at least 70% of the respondents regarded the dominant measures from state of the art (except IC.P.13.1 and IC.S.4.3) to be relevant for representing the capability of an agile team member.

On the whole, there were only three individual capability measures (*independent process and product audits* [IC.P.8.1], *software transition* [IC.P.9.1] and *perform software measurement process* [IC.P.12.2]) that were unfamiliar to at least 10% of the subjects and all of them were associated with professional category. In the case of these three measures, we also noticed a corresponding low response frequency for HR, suggesting that even respondents who were familiar with those found them to be less relevant.

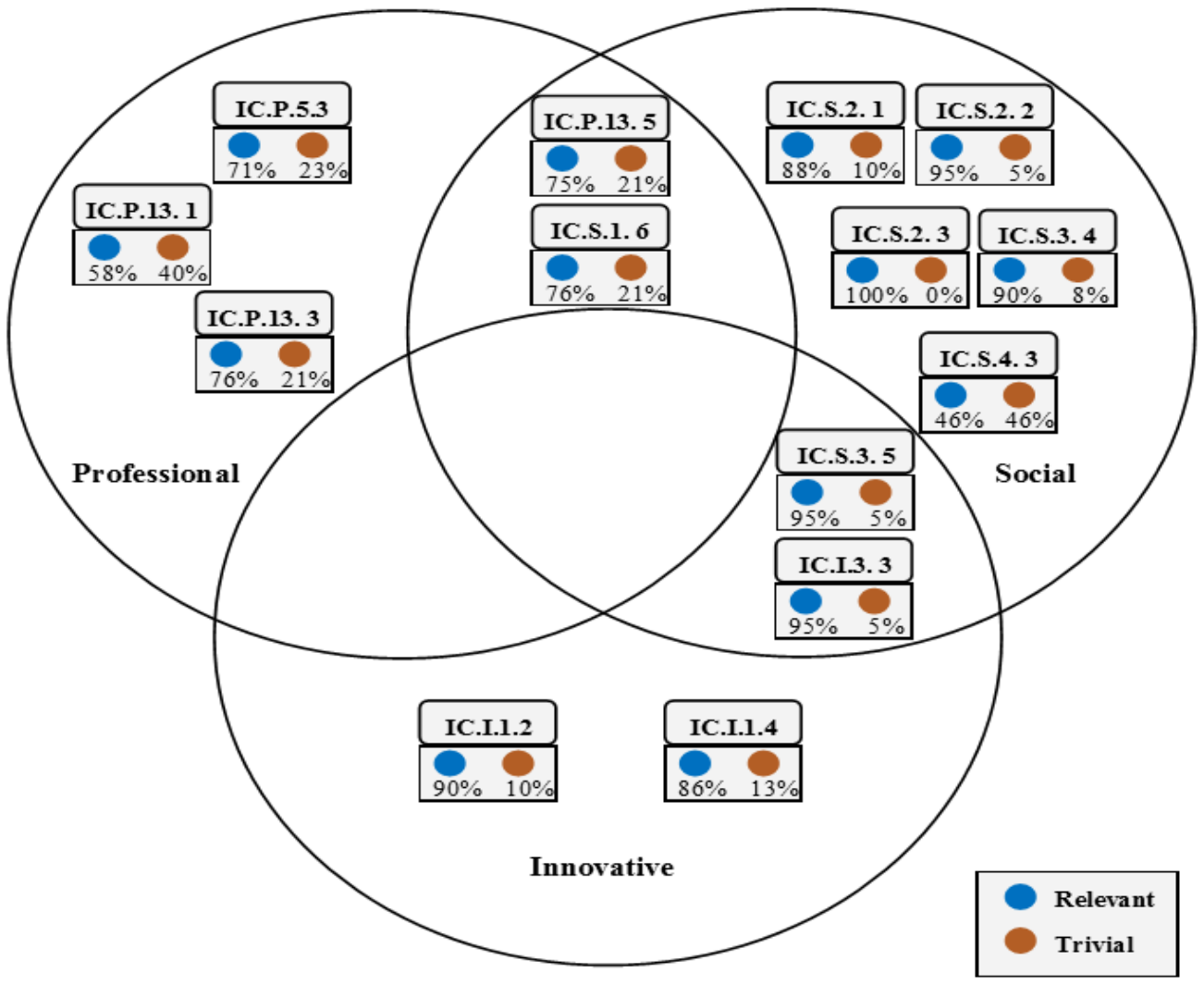

**Figure 3. Practitioners' perception of individual capability measures that were widely discussed in SE literature**

### 4.1.2. Perceived relevance of team capability measures (RQ.1.2)

We present heatmaps for displaying the responses in relation to each of the three primary categories of team capability measures. Further, within each primary category, we ranked and identified dominant capability measures by using FSC-1. The details pertaining to which capability measures were rated more relevant and how agile practitioners perceived the relevance of team level measures, are discussed next.

#### 4.1.2.1. Insights from the professional category

The heatmap from Table 7 displays the perceived relevance of team capability measures under the professional category.

- The top five measures which were indicated as HR by at least 43% of the respondents:
  - Growth (*active learning and improvement* [TC.P.5.1] and *advancement* [TC.P.5.2])
  - Business excellence (*result-orientation* [TC.P.3.2])
  - Team experience (*programming language experience* [TC.P.1.4])
  - *Clear goals* [TC.P.6.1]
- The list of measures that were not familiar to more than 5% of the respondents:
  - Agile capability (*conscious sensitivity* [TC.P.2.1])
  - Business excellence (*effectiveness* [TC.P.3.1])
  - *Team buy-in* [TC.P.6.3]
- Details of how agile practitioners perceived the relevance of widely discussed measures in SE literature [22] :
  - In relation to the experience of a team, *customer experience* [TC.P.1.1], *domain knowledge experience* [TC.P.1.2], *generational experience* [TC.P.1.3], *programming language experience* [TC.P.1.4] and *experience with tools* [TC.P.1.5] were discussed by two studies [44], [45]. The dominant measure among them was the *programming language experience* of a team. It was considered as relevant by 76% of the respondents, among whom, 43% regarded it as HR. Next, a team's skill in working with various software tools and project management tools (*experience with tools*) was regarded as relevant by around 76% of the respondents (HR=38%), followed by, the *domain knowledge experience* of a team, which was considered as relevant by around 78% of the respondents (HR=35%). A team's skill in seeing customer's perspective and acting accordingly (*customer experience*) was considered as relevant by around 56% of the respondents (HR=16%) and a team's experience with various phases of development of a software product (*generational experience*) was considered as relevant by around 63% of the respondents (HR=10%).

- The measure concerning expert skills or knowledge in a particular field possessed by a team (*expertise* [TC.P.6.4]) was discussed by two studies [44], [46]. This measure was considered as relevant by around 76% of the respondents (HR=36%).

**Table 7. Heatmap of response frequencies for team capability measures; Primary category: Professional. The IDs of measures with FSC-2 ranks are highlighted in bold**

| Sub-category | ID | Capability measure | Percentage | | | | | FSC-1 | FSC-2 |
|---|---|---|---|---|---|---|---|---|---|
| | | | *HR* | *R* | *SR* | *IR* | *IDK* | *Rank* | *Rank* |
| Team experience | TC.P.1.1 | Customer experience | 16.67 | 40 | 33.33 | 8.33 | 1.67 | #15 | |
| | TC.P.1.2 | Domain knowledge experience | 35 | 43.33 | 18.33 | 0 | 3.33 | #8 | |
| | TC.P.1.3 | Generational experience | 10 | 53.33 | 28.33 | 3.33 | 5 | #16 | |
| | **TC.P.1.4** | Programming language experience | 43.33 | 33.33 | 21.67 | 1.67 | 0 | #5 | #4 |
| | TC.P.1.5 | Experience with tools | 38.33 | 38.33 | 18.33 | 3.33 | 1.67 | #6 | |
| Agile capability | TC.P.2.1 | Conscious sensitivity | 20 | 50 | 18.33 | 3.33 | 8.33 | | |
| | TC.P.2.2 | Responsiveness to customer | 33.33 | 41.67 | 21.67 | 1.67 | 1.67 | #10 | |
| | TC.P.2.3 | Environment needs and changes | 33.33 | 40 | 18.33 | 3.33 | 5 | #11 | |
| Business excellence | TC.P.3.1 | Effectiveness | 43.33 | 45 | 3.33 | 1.67 | 6.67 | | |
| | TC.P.3.2 | Result-orientation | 55 | 33.33 | 8.33 | 0 | 3.33 | #2 | |
| | TC.P.3.3 | Systemic benefits | 30 | 51.67 | 13.33 | 0 | 5 | #12 | |
| Operational excellence | TC.P.4.1 | Sustainable efficiency | 33.33 | 46.67 | 15 | 0 | 5 | #9 | |
| | TC.P.4.2 | Consistent predictability | 20 | 61.67 | 10 | 3.33 | 5 | #14 | |
| Growth | **TC.P.5.1** | Active learning and improvement | 60 | 35 | 3.33 | 1.67 | 0 | #1 | #2 |
| | TC.P.5.2 | Advancement | 50 | 36.67 | 10 | 0 | 3.33 | #3 | |
| Miscellaneous | TC.P.6.1 | Clear goals | 50 | 36.67 | 11.67 | 0 | 1.67 | #4 | |
| | TC.P.6.2 | Full-time allocation | 25 | 38.33 | 30 | 5 | 1.67 | #13 | |
| | TC.P.6.3 | Team buy-in | 16.67 | 46.67 | 20 | 6.67 | 10 | | |
| | TC.P.6.4 | Expertise | 36.67 | 40 | 18.33 | 3.33 | 1.67 | #7 | |

#### 4.1.2.2. Insights from the social category

The heatmap from Table 8 displays the perceived relevance of team capability measures under the social category.

- The top five measures which were indicated as HR by at least 38% of the respondents:
  - *Cooperation* [TC.S.1.6]
  - *High motivation* [TC.S.1.2]
  - *Communication skills* [TC.S.1.5]
  - *Morale* [TC.S.1.1]
  - *Value diversity* [TC.S.1.3]
- *Cohesion* [TC.S.1.7], the degree to which team members want to contribute to the group in order to continue as a functioning work unit, was the only measure that was not familiar to more than 5% of the respondents.
- Details of how agile practitioners perceived the relevance of widely discussed measures in SE literature [22]:
  - The measure concerning the desire and energy in a team to be continually interested and committed towards attaining a goal (*high motivation* [TC.S.1.2]) was discussed by three studies [45], [46], [87]. It was considered as relevant by 90% of the respondents, among whom, 50% considered it to be HR.
  - A team's ability to work together (*cooperation* [TC.S.1.6]) was discussed by two studies [45], [88]. It was considered as relevant by 95% of respondents, among whom, 60% regarded it as HR.

**Table 8. Heatmap of response frequencies for team capability measures; Primary category: Social and innovative. The IDs of measures with FSC-2 ranks are highlighted in bold**

| Primary category | ID | Capability measure | Percentage | | | | | FSC-1 | FSC-2 |
|---|---|---|---|---|---|---|---|---|---|
| | | | *HR* | *R* | *SR* | *IR* | *IDK* | *Rank* | *Rank* |
| Social | TC.S.1.1 | Morale | 43.33 | 51.67 | 3.33 | 0 | 1.67 | #4 | |
| | TC.S.1.2 | High motivation | 50 | 40 | 6.67 | 1.67 | 1.67 | #2 | |
| | TC.S.1.3 | Value diversity | 38.33 | 45 | 8.33 | 5 | 3.33 | #5 | |
| | TC.S.1.4 | Internal competition | 15 | 41.67 | 26.67 | 15 | 1.67 | #6 | |
| | **TC.S.1.5** | Communication skills | 48.33 | 36.67 | 11.67 | 3.33 | 0 | #3 | #3 |
| | **TC.S.1.6** | Cooperation | 60 | 35 | 5 | 0 | 0 | #1 | #1 |
| | TC.S.1.7 | Cohesion | 31.67 | 50 | 11.67 | 0 | 6.67 | | |
| Innovative | **TC.I.1.1** | Creative exploration and exploitation | 31.67 | 53.33 | 11.67 | 3.33 | 0 | #2 | #5 |
| | TC.I.1.2 | Foresight | 35 | 51.67 | 11.67 | 0 | 1.67 | #1 | |

#### 4.1.2.3. Insights from the innovative category

The heatmap from Table 8 displays the perceived relevance of team capability measures under the innovative category.

- The measures *creative exploration and exploitation* [TC.I.1.1] and *foresight* [TC.I.1.2] were discussed by one study [38]. The dominant measure among them was *foresight*, which was considered as relevant by 86% of the respondents (HR=35%), followed by, *creative exploration and exploitation*, which was regarded as relevant by 85% of the respondents.

#### 4.1.2.4. Key insights from the three primary categories

Upon collating the capability measures from Table 7 and Table 8, and aggregating the percentages presented in the HR and R columns, we observed that all the 28 team capability measures were indicated as relevant by more than half of the respondents.

In order to explore which team capability measures were widely known as well as perceived as HR by majority of the respondents, we collectively analyzed the measures from the three primary categories by employing FSC-2. This resulted in identifying five measures, the ranks of which are presented under FSC-2 and the IDs are highlighted in bold. These five measures were perceived as highly relevant for characterizing the capability of an agile team, by at least 31% of the respondents. Respondents considered the following five measures as highly appropriate for characterizing the capability of an agile team:

- The ability to work together (*cooperation*)
- The ability to engage in reading, writing, talking, listening and reflecting, with an intention of getting better (*active learning and improvement*)
- The ability to convey or share ideas and feelings effectively (*communication skills*)
- Proficiency in working with different programming languages (*programming language experience*)
- The ability to use a creative process to explore feelings, ideas and questions, together with making use of and benefiting from resources (*creative exploration and exploitation*)

A Venn diagram (see Figure 4) presents a summary of how agile practitioners perceived the relevance of team capability measures, especially the ones that were widely discussed in SE literature. From Figure 4, we can clearly see that more than 75% of the respondents regarded the dominant measures from state of the art (except TC.P.1.1 and TC.P.1.3) to be relevant for representing the capability of an agile team.

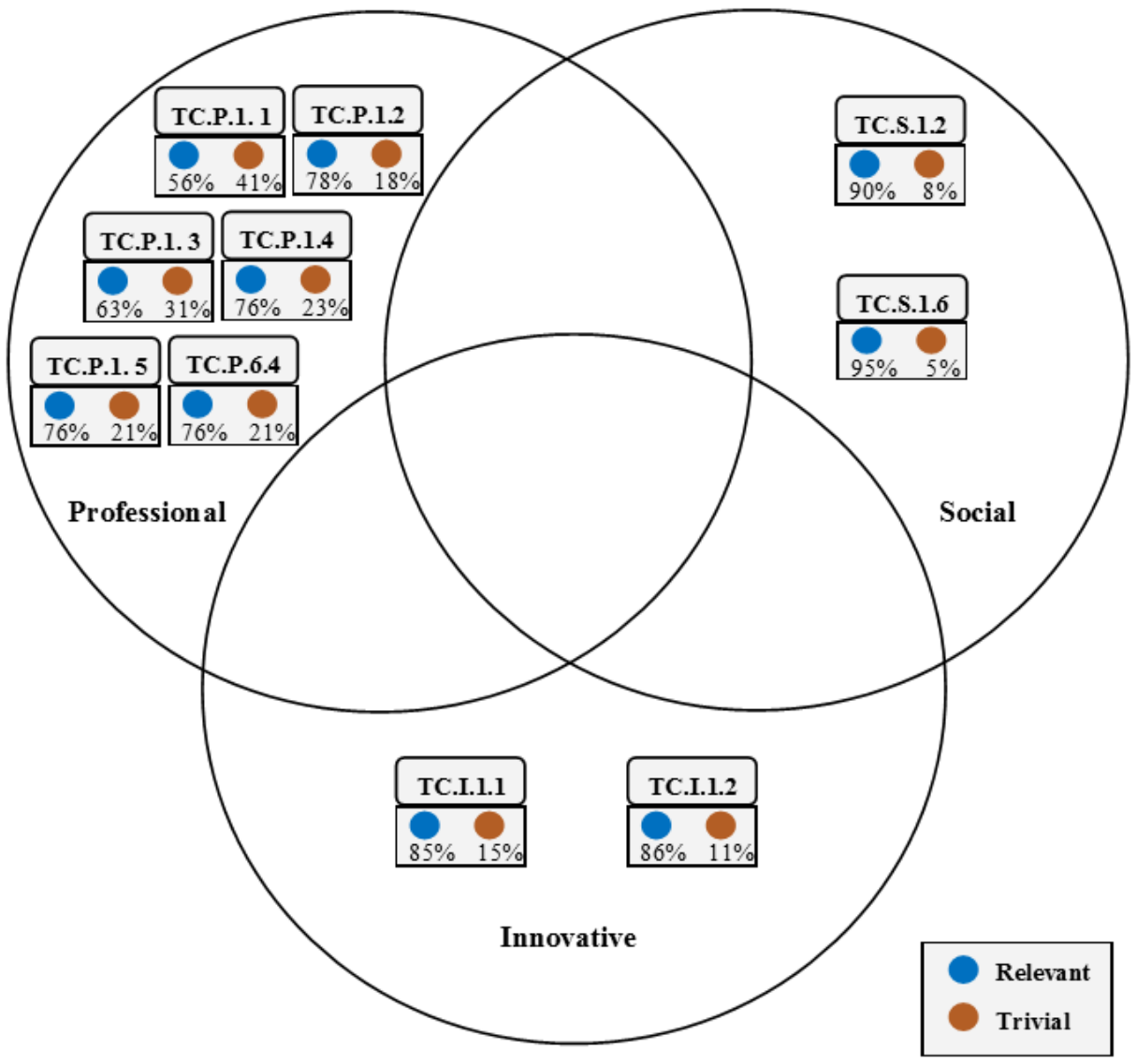

**Figure 4. Practitioners' perception of team capability measures that were widely discussed in SE literature**

## 4.2. Examining the perceptions across sub-groups (RQ.2)

The diversified sample acquired by our survey, where experienced professionals held a range of roles and worked with various software development methodologies, enabled us to examine how the perceived relevance of capability measures vary across different groups of practitioners. The responses to demographic questions were used to create different sub-groups, which were employed towards analyzing the variations in the perceptions, as presented next.

### 4.2.1. Differences in the perceptions across sub-groups (RQ.2.1)

Differences in the ratings of the capability measures across different sub-groups were analyzed by means of the Kruskal–Wallis test (K-W test), a nonparametric statistical method that compares independent groups of sample data. To facilitate the analysis, respondents' demographic data was used to segregate the survey responses into sub-groups. Since the K-W test requires the sample size of each group to be at least five [89], the sub-groups with too few observations (less than five) were omitted from this test.

Our null hypothesis for K-W test is, there will not be a difference between the perceptions of groups of respondents in relation to capability measures assessment. In cases where the null-hypothesis was rejected and a significant difference was noticed, the Dunn's test of multiple comparisons based on rank sums with Bonferroni correction [90] was used to examine which of the sub-groups significantly (significance level was set to 0.05) differed from the rest.

We further computed epsilon square ($\varepsilon^2$) to report the effect size [91] whenever the grouping variable has an effect on the ordinal-scale variable under observation. According to Rea and Parker [92], the following are the interpretations for $\varepsilon^2$ value: weak effect ($0.01 < \varepsilon^2 < 0.04$), moderate ($0.04 < \varepsilon^2 < 0.16$), relatively strong ($0.16 < \varepsilon^2 < 0.36$), strong ($0.36 < \varepsilon^2 < 0.64$) and very strong ($0.64 < \varepsilon^2 < 1$).

#### 4.2.1.1. Grouping criterion 1: Respondents' organizational domain

On the basis of the details of respondents' organizational domain(s) from Table 2, we identified two domains (ICT and Web applications) with at least five observations. The rest of the observations were grouped under 'other' category, thus, effectively leading to three sub-groups: ICT (n= 31), Web applications (n= 5) and other (n= 24).

By employing a series of K-W tests, we identified two individual capability measures where the organizational domain of the respondents had a significant ($p < .05$) moderate effect on their perceptions (see Table 9. Note: Individual Capability (IC) measures can be distinguished from the Team Capability (TC) measures by ID). Post-hoc tests using Dunn's test with Bonferroni correction showed significant differences between ICT group and 'other' group ($p < .05$). Respondents associated with ICT domain perceived *feeling* [IC.S.4.7] to be significantly more relevant than 'other' group. Whereas in the case of *intelligent risk taking* [IC.I.5.2], respondents from 'other' group perceived the measure to be more relevant than ICT group.

Next, K-W tests in relation to team capability measures showed that organizational domain had a significant moderate effect on how practitioners perceived operational excellence measures (see Table 9). Post-hoc tests showed that members from Web applications group considered *sustainable efficiency* [TC.P.4.1] to be significantly more relevant than ICT group members. Whereas in the case of *consistent predictability* [TC.P.4.2], Web applications group perceived the measure to be more relevant than the 'other' group.

**Table 9. Results of K-W test: cases where organizational domain had a significant effect on responses**

| Sub-category | ID | Capability measure | p – value (K-W test) | H - statistic | Epsilon square ($\varepsilon^2$) | Groups that significantly differ | p - value (Dunn's test) |
|---|---|---|---|---|---|---|---|
| Personal | IC.S.4.7 | Feeling | 0.019 | 7.912 | 0.134 | ICT - Other | 0.015 |
| Managing change | IC.I.5.2 | Intelligent risk-taking | 0.020 | 7.779 | 0.131 | ICT - Other | 0.018 |
| Operational excellence | TC.P.4.1 | Sustainable efficiency | 0.020 | 7.797 | 0.132 | ICT – Web applications | 0.017 |
| | TC.P.4.2 | Consistent predictability | 0.020 | 7.865 | 0.133 | Web applications - Other | 0.015 |

#### 4.2.1.2. Grouping criterion 2: ASD methodology adopted by respondents' team

We inspected the details of the ASD methodologies used at respondents' organizations (see Table 3) and segregated their responses into three sub-groups: Scrum (n=41), Kanban (n=8) and Scrumban (n=5). Upon performing a sequence of K-W tests, in the case of eight individual capability measures, we observed that the ASD methodology employed by respondents had a significant moderate effect on their perceptions (see Table 10). More importantly, we observed that ASD methodology had a relatively strong effect on how respondents perceived the measure concerning proficiency in developing software release plan (*manage software releases* [IC.P.11.3]). Post-hoc tests conducted in relation to the nine measures (including IC.P.11.3) showed that the respondents adhering to Scrumban perceived the measure *feeling* [IC.S.4.7] to be significantly more relevant than the Scrum group and the respondents adhering to Kanban perceived the rest of the eight measures to be significantly more relevant than the Scrum group.

A series of K-W tests for team capability measures showed that ASD methodology had a relatively strong effect on how practitioners perceived the relevance of team level *communication skills* [TC.S.1.5]. A post-hoc test showed that the Kanban group members regarded *communication skills* as significantly more relevant than the Scrum group.

**Table 10. Results of K-W test: cases where ASD methodology had a significant effect on responses**

| Sub-category | ID | Capability measure | p – value (K-W test) | H - statistic | Epsilon square ($\varepsilon^2$) | Groups that significantly differ | p - value (Dunn's test) |
|---|---|---|---|---|---|---|---|
| Software security and safety | IC.P.10.3 | Quality | 0.044 | 6.255 | 0.118 | Scrum - Kanban | 0.037 |
| Software configuration management | IC.P.11.3 | Manage software releases | 0.002 | 12.058 | 0.227 | Scrum - Kanban | 0.002 |
| Miscellaneous | IC.P.13.1 | Prior work experience | 0.017 | 8.140 | 0.153 | Scrum - Kanban | 0.036 |
| | IC.P.13.5 | Planning skills | 0.047 | 6.133 | 0.115 | Scrum - Kanban | 0.046 |
| Personal | IC.S.4.7 | Feeling | 0.037 | 6.608 | 0.124 | Scrum – Scrumban | 0.034 |
| | IC.S.4.20 | Desire to improve things | 0.019 | 7.950 | 0.15 | Scrum - Kanban | 0.026 |
| Creativity | IC.I.1.1 | Generating ideas | 0.026 | 7.265 | 0.137 | Scrum - Kanban | 0.026 |
| Enterprising | IC.I.2.3 | Gathering and evaluating information | 0.022 | 7.594 | 0.143 | Scrum - Kanban | 0.018 |
| Integrating perspectives | IC.I.3.4 | Engaging in non-work related interests | 0.049 | 6.037 | 0.114 | Scrum - Kanban | 0.049 |
| Social | TC.S.1.5 | Communication skills | 0.009 | 9.535 | 0.180 | Scrum - Kanban | 0.008 |

#### 4.2.1.3. Grouping criterion 3: Respondents' primary role

Looking into the details of the respondents' primary role, we segregated their responses into four sub-groups: CI engineer (n=5), Scrum master (n=5), developer (n=41) and tester (n=5). A series of K-W tests indicated that the respondents' primary role had a significant relatively strong effect (see Table 11) on how they regarded the *collaborating* [IC.I.3.3] measure. Interestingly, all the CI engineers regarded the collaborating measure to be relevant and all the Scrum masters perceived it to be highly relevant. Whereas in the case of responses for *managing the future* [IC.I.4.3] measure, the primary role of the respondents had a significant moderate effect. The Scrum masters regarded it as significantly more relevant than the CI engineers. Further, we observed that the respondents' primary role had a relatively strong effect on how they perceived a team's *experience with programming languages* [TC.P.1.4]. A post-hoc test showed that Scrum masters regarded the *programming language experience* as significantly more relevant than CI engineers.

**Table 11. Results of K-W test: cases where primary role had a significant effect on responses**

| Sub-category | ID | Capability measure | p – value (K-W test) | H - statistic | Epsilon square ($\varepsilon^2$) | Groups that significantly differ | p - value (Dunn's test) |
|---|---|---|---|---|---|---|---|
| Integrating perspectives | IC.I.3.3 | Collaborating | 0.016 | 10.282 | 0.194 | CI engineer – Scrum master | 0.030 |
| Forecasting | IC.I.4.3 | Managing the future | 0.039 | 8.375 | 0.158 | CI engineer – Scrum master | 0.044 |
| Team experience | TC.P.1.4 | Programming language experience | 0.032 | 8.800 | 0.166 | CI engineer – Scrum master | 0.040 |

#### 4.2.1.4. Grouping criterion 4: Respondents' work experience

We inspected the details of work experience reported by the respondents (see Figure 2.2) and segregated their responses into five sub-groups: three or less than four [3, 4) years (n=6), four or less than five [4, 5) years (n=21), five or less than six [5, 6) years (n=21), six or less than seven [6, 7) years (n=5) and more than 7 years (n=6). By employing a series of K-W tests, we identified four individual capability measures where respondents' work experience had a significant relatively strong effect on their perceptions (see Table 12).

Post-hoc tests conducted in relation to the human-computer interaction measures showed that the respondents who had six or less than seven years of experience perceived the *interaction style design* [IC.P.6.2] and *visual design* [IC.P.6.3] measures to be significantly more relevant than respondents who had [3, 4) years of experience. Post-hoc tests conducted in relation to personal sub-category measures showed that the respondents who had [5, 6) years of experience considered the measure *thinking* [IC.S.4.6] to be significantly more relevant than respondents who had [3, 4) years of experience. Similarly, in the case of *maintaining big picture view* [IC.S.4.22], respondents who had [5, 6) years of experience considered the measure to be significantly more relevant than respondents who had [4, 5) years of experience. With respect to the aforementioned measures, among the sub-groups where significant differences were noticed, we observed that the perceived relevance was directly proportional to the experience of the group.

Additionally, respondents' work experience had a relatively strong effect on how they perceived a team's ability to judge future events (*foresight* [TC.I.1.2]). A post-hoc test showed that respondents with [6, 7) years of experience considered *foresight* to be significantly more relevant than respondents who had more than seven years of experience.

**Table 12. Results of K-W test: cases where work experience had a significant effect on responses**

| Sub-category | ID | Capability measure | p – value (K-W test) | H - statistic | Epsilon square ($\varepsilon^2$) | Groups that significantly differ | p - value (Dunn's test) |
|---|---|---|---|---|---|---|---|
| Human-computer interaction | IC.P.6.2 | Interaction style design | 0.022 | 11.396 | 0.20 | [3, 4) – [6, 7) | 0.011 |
| | IC.P.6.3 | Visual design | 0.008 | 13.753 | 0.241 | [3, 4) – [6, 7) | 0.006 |
| Personal | IC.S.4.6 | Thinking | 0.011 | 13.026 | 0.228 | [3, 4) – [5, 6) | 0.028 |
| | IC.S.4.22 | Maintaining 'big picture' view | 0.021 | 11.597 | 0.203 | [4, 5) – [5, 6) | 0.017 |
| Innovative | TC.I.1.2 | Foresight | 0.033 | 10.452 | 0.183 | [6, 7) – > 7 | 0.042 |

#### 4.2.1.5. Grouping criterion 5: Respondents' team size

We used the details of respondents' team size, reported in Figure 2.1, to segregate the responses into three sub-groups: less than or equal to five [1, 5] members (n=16), six to 10 [6, 10] members (n=33) and 11 to 15 [11, 15] members (n=8). Upon performing a sequence of K-W tests, in the case of four individual capability measures, we observed that respondents' team size had a significant moderate effect on their perceptions (see Table 13).

A post-hoc test conducted in relation to the *software requirements specification* proficiency [IC.P.1.3] showed that the respondents whose teams consisted of [10, 15] members perceived the measure to be significantly more relevant than respondents whose team had [1, 5] members. Further, in relation to the other three individual capability measures (IC.P.3.3, IC.S.1.4 and IC.S.2.3), the respondents whose teams consisted of [6, 10] members perceived the measures to be significantly more relevant than respondents whose team size was less than five.

Further, we observed that the respondents' team size had a significant moderate effect on how they perceived the team level capability measures: *conscious sensitivity* [TC.P.2.1] and *value diversity* [TC.S.1.3]. A post-hoc test showed that respondents whose teams consisted of [6, 10] members perceived the *value diversity* to be significantly more relevant than the respondents whose teams had [1, 5] members. Similarly, the respondents whose teams had [10, 15] members perceived *conscious sensitivity* to be significantly more relevant than respondents whose team size was less than five.

**Table 13. Results of K-W test: cases where team size had a significant effect on responses**

| Sub-category | ID | Capability measure | p – value (K-W test) | H - statistic | Epsilon square ($\varepsilon^2$) | Groups that significantly differ | p - value (Dunn's test) |
|---|---|---|---|---|---|---|---|
| Software requirements | IC.P.1.3 | Specification | 0.047 | 6.110 | 0.109 | [1, 5] - [10, 15] | 0.049 |
| Software system engineering | IC.P.3.3 | Software-intensive systems engineering | 0.034 | 6.743 | 0.120 | [1, 5] - [6, 10] | 0.028 |
| Affective | IC.S.1.4 | Work ethic | 0.032 | 6.856 | 0.122 | [1, 5] - [6, 10] | 0.027 |
| Communication | IC.S.2.3 | Questioning skills | 0.030 | 7.040 | 0.125 | [1, 5] - [6, 10] | 0.032 |
| Agile capability | TC.P.2.1 | Conscious sensitivity | 0.048 | 6.080 | 0.108 | [1, 5] - [10, 15] | 0.041 |
| Social | TC.S.1.3 | Value diversity | 0.048 | 6.092 | 0.108 | [1, 5] - [6, 10] | 0.048 |

#### 4.2.1.6. Key insights from the sub-groups

Criteria such as ASD methodology, primary role and work experience had a relatively strong effect on how respondents perceived some capability measures. However, when comparing the results among the aforementioned three criteria, we noticed that respondents' work experience was the one that had a significant relatively strong effect on the perceptions of the majority of the capability measures (four individual capability measures and one team capability measure). However, since some sub-groups in our study comprised a small sample, we cannot generalize our findings without further investigation. With respect to the contribution of our findings to research, the strong effect sizes observed in our study indicate that the differences between sub-groups are highly likely to be observed in other investigations that will be executed within an ASD context.

### 4.2.2. Predominant capability measures across sub-groups (RQ.2.2)

Here, we examine the perceptions within sub-groups for figuring out which measures were widely indicated as relevant for characterizing the capability of an individual or a team. We resorted to the same sub-groups listed in Section 4.2.1 (RQ.2.1) for answering this question. In order to facilitate our analysis within sub-groups, for each capability measure, we calculate the proportion of responses that were indicated as highly relevant (best response), relevant (positive response), trivial (negative response) and highly trivial (worst response).

More formally, let *n(H), n(R), n(SR)* and *n(IR)* denote the number of HR, R, SR and IR responses received for a capability measure within a group. We calculated the below-mentioned statistics [76] for different sub-groups and determined the highest rated measures within each group, by sorting all the capability measures in terms of their HR-scores (descending), followed by, R-scores (descending), T-scores (ascending) and HT-scores (ascending):

**HR-score:** The percentage of ratings (all ratings excluding IDK) that were 'highly relevant'

$$\text{HR-score} = \frac{n(H)}{n(H)+n(R)+n(SR)+n(IR)} \qquad (1)$$

**R-score:** The percentage of ratings that were 'relevant'

$$\text{R-score} = \frac{n(H)+n(R)}{n(H)+n(R)+n(SR)+n(IR)} \qquad (2)$$

**T-score:** The percentage of ratings that were 'trivial'

$$\text{T-score} = \frac{n(SR)+n(IR)}{n(H)+n(R)+n(SR)+n(IR)} \qquad (3)$$

**HT-score:** The percentage of ratings that were 'highly trivial'

$$\text{HT-score} = \frac{n(IR)}{n(H)+n(R)+n(SR)+n(IR)} \qquad (4)$$

The HR scores and R scores of individual and team capability measures were further visually analyzed using boxplots. Boxplots are useful in comparing the shape of data when separated by a categorical variable. Besides presenting the median value of a distribution, boxplots also show whether there are any extreme observations (outliers). The box and whiskers represent the 25–75th and 10–90th percentiles (non-outlier range), respectively. These plots help in visualizing the spread (variance) of data across different categories.

#### 4.2.2.1. Grouping criterion 1: Respondents' organizational domain

Within the three sub-groups created based on the respondents' organizational domain, the set of HR-scores and R-scores calculated for each capability measure were used to generate boxplots (see Figure 5). Across the three sub-groups, although the mid-points of the HR-scores for individual capability measures seem to be relatively close to each other, the variations in the sizes of the boxes and whiskers indicate different distributions of views among members. In specific, the sizes of the box and upper whisker for Web-applications sub-group appear to be bigger in comparison to the rest of the sub-groups. This informs us that, among the members from the Web-applications sub-group, there were greater disparities in the individual capability measures reported as HR. In the case of team capability measures, the size of the boxplot for ICT sub-group appears to be smaller than the corresponding plot for individual capability measures. This indicates higher level of conformity among the team capability measures (rather than individual level measures) reported as HR and R, by members from ICT domain.

In the case of Web applications sub-group, except for nine individual capability measures (R-score = 0.33 [IC.S.4.3, IC.S.4.7, IC.I.2.5, IC.S.4.8, IC.S.4.27, IC.P.3.6, IC.P.13.2], R-score = 0.16 [IC.S.4.13] and R-score = 0 [IC.S.4.2]), the R-scores for the rest of the individual and team capability measures were observed to be at least 0.50. Moreover, upon comparing the distribution of R-scores across the three sub-groups, the median for the Web applications sub-group appears to be higher than the other two groups. This indicates that members associated with the Web applications domain rated greater proportion of the capability measures as relevant.

The scores of capability measures in each sub-group were further sorted and a list of top measures (both individual and team level) was prepared. Due to space constraints, we only present and discuss the top five capability measures from each sub-group. These measures are presented in Table 14, where the ones that were common across all the sub-groups are highlighted in bold. The key observations from Table 14 are presented as follows:

- The HR-score for *responsibility* [IC.S.5.5] across the three sub-groups was observed to be at least 0.70. This means, within each of the three sub-groups, at least 70% of the ratings received for the measure indicate it to be highly relevant for representing individual capability.
- At least 64% of the ratings from members of ICT group and other group indicate the following social context measures to be highly relevant for representing individual capability:
    - Communication (*listening skills* [IC.S.2.2])
    - Work ethics (*motivation to work* [IC.S.5.3])
- In relation to individual capability measures, members from the ICT group and Web applications group commonly regarded *commitment* [IC.S.5.4] to be highly relevant. Whereas members from the Web applications group and other group commonly considered the measure *teamwork oriented* [IC.S.3.5] to be highly relevant.
- In relation to the team capability measures, the top five measures identified from the Web applications group differed from the other two sub-groups.
- Members from ICT group and other group commonly considered the following measures to be highly relevant for representing team capability:
    - Growth (*active learning and improvement* [TC.P.5.1])
    - *Cooperation* [TC.S.1.6] and *high motivation* [TC.S.1.2]
    - Business excellence (*result-orientation* [TC.P.3.2])

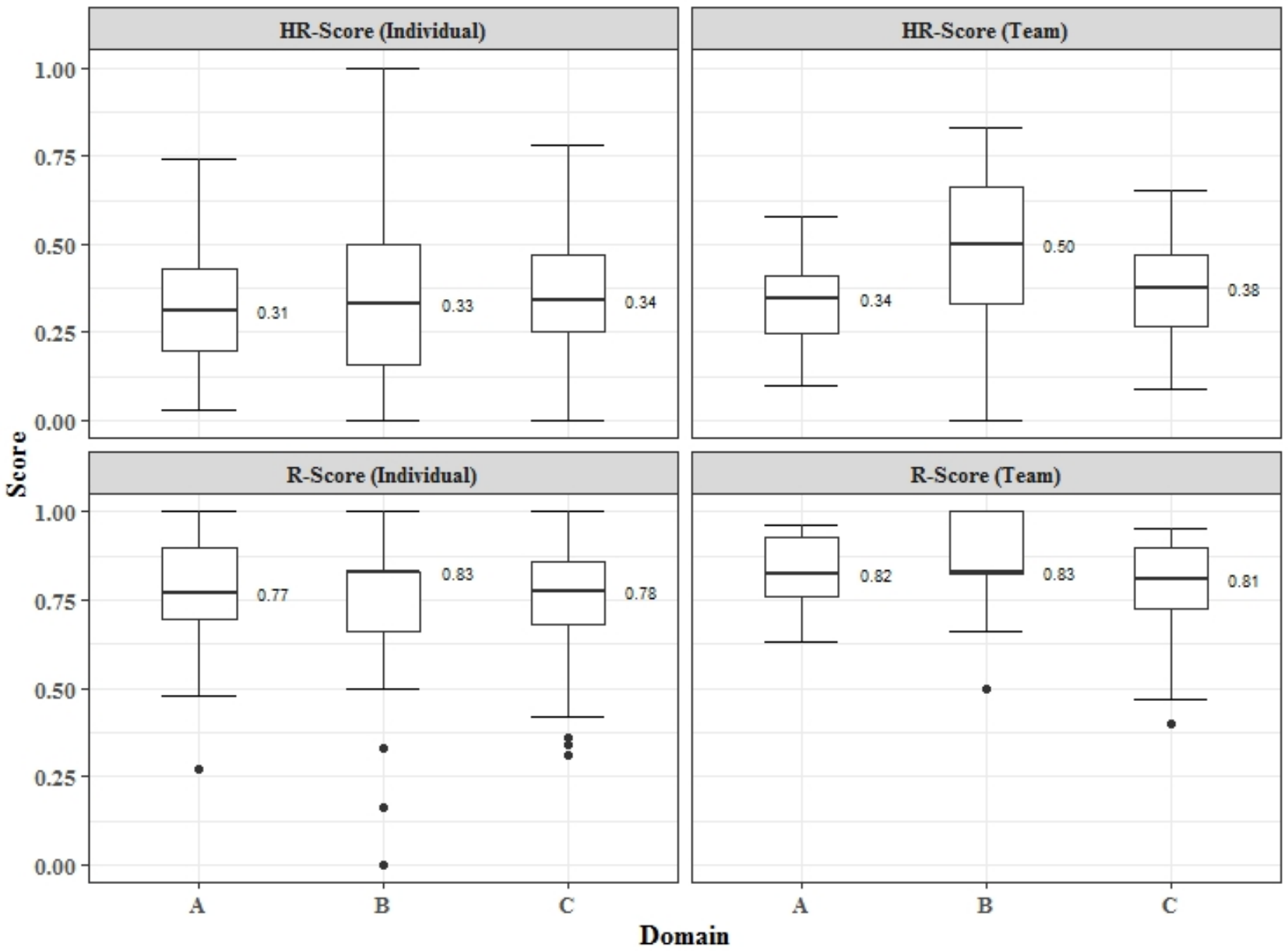

**Figure 5. Distribution of scores for capability measures; Grouping criterion: Organizational domain (A: ICT, B: Web applications & C: Other)**

**Table 14.Top five rated individual and team capability measures; Grouping criterion: Organizational domain. The IDs of measures that are common across all the sub-groups are highlighted in bold**

| Type | ICT | | | Web applications | | | Other | | |
|---|---|---|---|---|---|---|---|---|---|
| | ID | HR-score | R-score | ID | HR-score | R-score | ID | HR-score | R-score |
| Individual capability measures | IC.S.2.2 | 0.74 | 0.93 | IC.S.3.4 | 1 | 1 | **IC.S.5.5** | 0.78 | 1 |
| | **IC.S.5.5** | 0.7 | 1 | IC.S.1.4 | 0.83 | 1 | IC.S.5.3 | 0.77 | 0.95 |
| | IC.S.2.3 | 0.64 | 1 | IC.S.3.5 | 0.83 | 1 | IC.S.1.9 | 0.69 | 1 |
| | IC.S.5.4 | 0.64 | 1 | IC.S.5.4 | 0.83 | 1 | IC.S.2.2 | 0.69 | 0.95 |
| | IC.S.5.3 | 0.64 | 0.93 | **IC.S.5.5** | 0.83 | 1 | IC.S.3.5 | 0.69 | 0.86 |
| Team capability measures | TC.P.5.1 | 0.58 | 0.93 | TC.P.4.1 | 0.83 | 1 | TC.P.5.1 | 0.65 | 0.95 |
| | TC.S.1.6 | 0.58 | 0.93 | TC.S.1.5 | 0.83 | 1 | TC.P.3.2 | 0.63 | 0.86 |
| | TC.P.6.1 | 0.54 | 0.9 | TC.P.1.4 | 0.83 | 0.83 | TC.S.1.6 | 0.6 | 0.95 |
| | TC.S.1.2 | 0.51 | 0.93 | TC.P.2.2 | 0.66 | 1 | TC.P.5.2 | 0.54 | 0.9 |
| | TC.P.3.2 | 0.5 | 0.93 | TC.P.3.1 | 0.66 | 1 | TC.S.1.2 | 0.54 | 0.9 |

#### 4.2.2.2. Grouping criterion 2: ASD methodology adopted by respondents' team

Among the three sub-groups created based on the ASD methodology adopted at respondents' organization, the set of HR-scores and R-scores calculated for each capability measure were used to generate boxplots (see Figure 6). In the case of Kanban and Scrumban sub-groups, we can notice greater dispersions in the HR-scores for individual capability measures. The elongated lower whisker in the case of Kanban sub-group and the relatively bigger size box in the case of Scrumban sub-group, informs us that there were greater disparities in the HR-scores of at least 25% (Kanban sub-group) and 50% (Scrumban sub-group) of the individual capability measures.

On the other hand, the size of the HR-score boxplot (individual capability measures) for Scrum sub-group appears to be relatively smaller than other sub-groups. The small size of the boxplot indicates higher level of agreement among the individual measures reported as HR, by members practicing Scrum. In the case of team capability measures, the size of the boxplot for Scrumban sub-group appears to be smaller than other sub-groups and also the corresponding plot for individual measures. This indicates that, in Scrumban sub-group, there is high level of conformity among the team capability measures reported as HR and R. Upon comparing the distribution of HR-scores (individual and team measures) across the three sub-groups, the median for the Kanban sub-group appears to be higher than the other two groups. This shows members practicing Kanban rated greater proportion of the measures as highly relevant.

Whereas when the distributions of R-scores are compared, within the Scrumban sub-group, we can clearly see that the R-score was 1 for almost 50% of the individual and team capability measures. This indicates that all the members from the Scrumban sub-group considered around half of the capability measures as relevant. Since the median of R-scores for the Scrumban sub-group appears to be higher than the other two groups, it is further evident that members adhering to Scrumban methodology rated greater proportion of capability measures as relevant.

The top measures from the three sub-groups are presented in Table 15 and the key observations are as follows:

- Within each of the three sub-groups, at least 68% of the ratings received for *listening skills* [IC.S.2.2] indicate it to be highly relevant for representing individual capability.
- Members adhering to Scrum and Kanban practices commonly considered *responsibility* [IC.S.5.5] to be highly relevant for representing individual capability.
- Within each of the three sub-groups, at least 58% of the ratings received for *cooperation* [TC.S.1.6] and *active learning and improvement* [TC.P.5.1] indicate the measures to be highly relevant for representing team capability.
- Members adhering to Scrum and Scrumban practices commonly considered the measure *advancement* [TC.P.5.2] to be highly relevant for representing team capability.

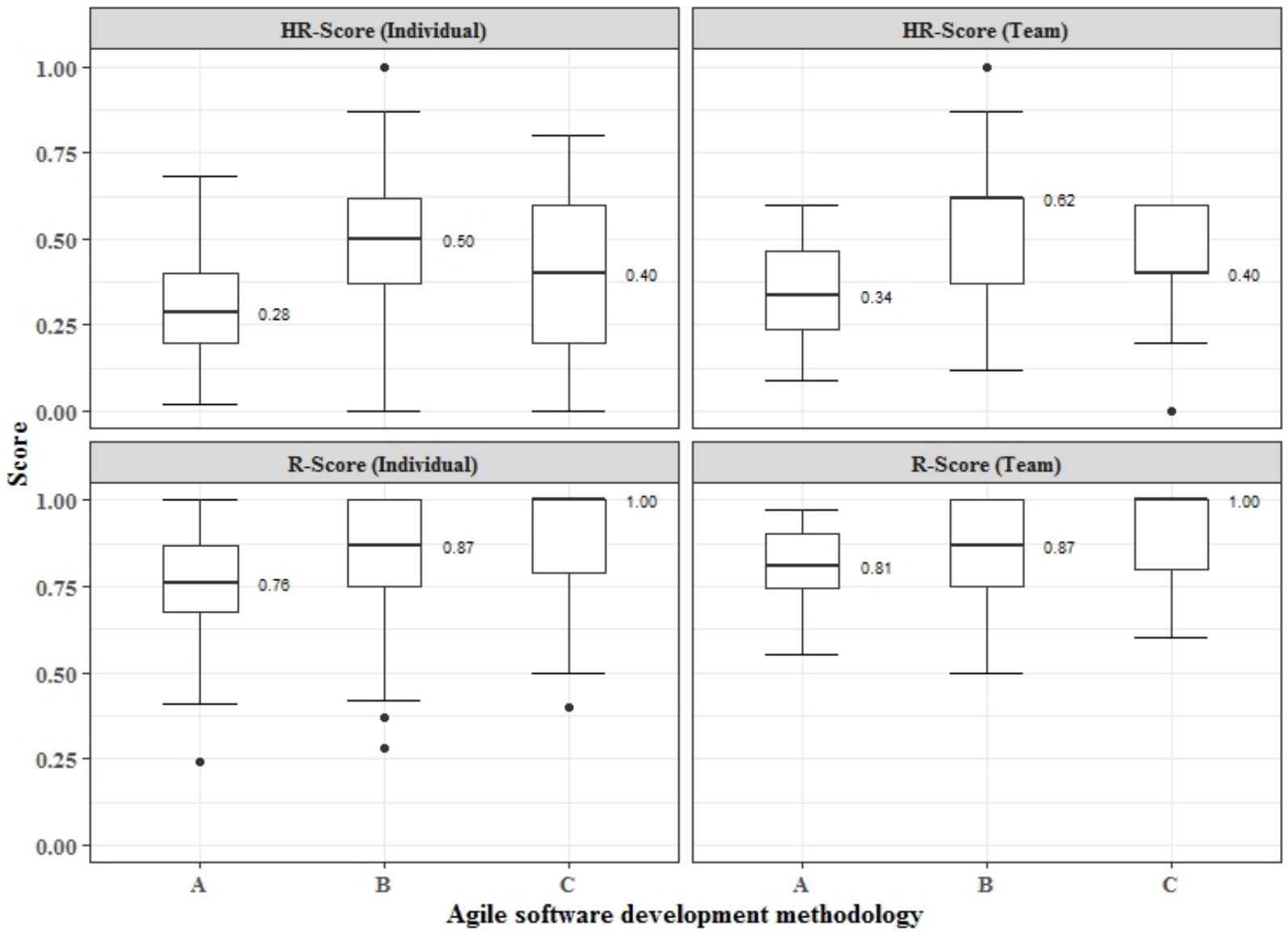

**Figure 6. Distribution of scores for capability measures; Grouping criterion: ASD methodology (A: Scrum, B: Kanban & C: Scrumban)**

**Table 15. Top five rated individual and team capability measures; Grouping criterion: ASD methodology. The IDs of measures that are common across all the sub-groups are highlighted in bold**

| Type | Scrum | | | Kanban | | | Scrumban | | |
|---|---|---|---|---|---|---|---|---|---|
| | ID | HR-score | R-score | ID | HR-score | R-score | ID | HR-score | R-score |
| Individual capability measures | IC.S.5.5 | 0.68 | 1 | IC.S.4.20 | 1 | 1 | **IC.S.2.2** | 0.8 | 1 |
| | **IC.S.2.2** | 0.68 | 0.92 | IC.S.5.5 | 1 | 1 | IC.S.3.1 | 0.8 | 1 |
| | IC.S.3.5 | 0.65 | 0.97 | IC.I.2.3 | 1 | 1 | IC.S.4.1 | 0.8 | 1 |

| | Scrum | | | Kanban | | | Scrumban | | |
|---|---|---|---|---|---|---|---|---|---|
| | **ID** | **HR-score** | **R-score** | **ID** | **HR-score** | **R-score** | **ID** | **HR-score** | **R-score** |
| | IC.S.5.3 | 0.65 | 0.9 | IC.P.10.3 | 0.87 | 1 | IC.S.4.6 | 0.8 | 1 |
| | IC.S.2.3 | 0.63 | 1 | **IC.S.2.2** | 0.87 | 1 | IC.S.4.15 | 0.8 | 1 |
| Team capability measures | **TC.P.5.1** | 0.6 | 0.92 | TC.S.1.5 | 1 | 1 | TC.P.4.2 | 0.6 | 1 |
| | **TC.S.1.6** | 0.58 | 0.95 | **TC.S.1.6** | 0.87 | 1 | **TC.P.5.1** | 0.6 | 1 |
| | TC.P.6.1 | 0.57 | 0.92 | TC.I.1.2 | 0.85 | 1 | TC.P.5.2 | 0.6 | 1 |
| | TC.P.3.2 | 0.56 | 0.97 | **TC.P.5.1** | 0.75 | 1 | TC.P.6.4 | 0.6 | 1 |
| | TC.P.5.2 | 0.53 | 0.92 | TC.P.3.1 | 0.75 | 0.87 | **TC.S.1.6** | 0.6 | 1 |

#### 4.2.2.3. Grouping criterion 3: Respondents' primary role

Among the four sub-groups created based on the primary role of the respondents, the set of HR-scores and R-scores calculated for each capability measure were used to generate boxplots (see Figure 7). Looking at the distribution of HR-scores (individual capability measures) for Scrum master sub-group, it is evident that around half of the individual capability measures (quartile group three and four) were regarded as highly relevant by more than 50% of the members. Upon comparing the distribution of HR-scores (both individual and team level measures) across the four sub-groups, the median for the Scrum master sub-group appears to be higher than the other three groups. This indicates that the Scrum masters perceived greater proportion of capability measures as highly relevant.

Upon comparing the distribution of R-scores, the median for the CI engineer sub-group (individual capability measures) can be seen to be higher, indicating that the CI engineers perceived greater proportion of individual capability measures as relevant. Within the Scrum master sub-group, we can clearly see that the R-score was 1 for almost 50% of the team capability measures. This indicates that all the Scrum masters in our study sample unanimously considered around half of the team capability measures as relevant.

In the case of the boxplots for developer sub-group, the interquartile range appears to be relatively smaller than that of the other three sub-groups. This indicates that, for at least 50% of the capability measures (both individual and team level measures), there is a high level of conformity among the developers' perceptions.

The list of top measures from each sub-group is presented in Table 16 and the key aspects from the table are as follows:

- At least 67% of the ratings from CI engineer group and developer group indicate the following measures to be highly relevant for representing individual capability:
  - Work ethics (*motivation to work* [IC.S.5.3])
  - Enterprising (*seeking improvement* [IC.I.2.2])
- Among CI engineers, Scrum masters and developers, at least 70% of the ratings received for *responsibility* [IC.S.5.5] indicate it to be highly relevant for representing individual capability.
- In relation to individual capability measures, members from CI engineer group and Scrum master group commonly regarded *team participation skills* [IC.S.1.9] to be highly relevant. Whereas members from developer and tester sub-groups commonly considered the measure *questioning skills* [IC.S.2.3] to be highly relevant.
- At least 50% of the ratings from Scrum master group and tester group indicate the following measures to be highly relevant for representing team capability:
  - Business excellence (*effectiveness* [TC.P.3.1])
  - Growth (*advancement* [TC.P.5.2])
- Among the top five ranked measures from CI engineers and developers, the measure *cooperation* [TC.S.1.6] was observed to be common, where at least 60% of the ratings received for the measure regarded it to be highly relevant. Whereas among CI engineers and testers, at least 50% of the ratings received for the measure *morale* [TC.S.1.1] indicate it to be highly relevant for representing team capability.

- Among the top five ranked measures from developers and testers, the measure *clear goals* [TC.P.6.1] was observed to be common, where at least 50% of the ratings received for the measure indicate it to be highly relevant. Whereas among developers and Scrum masters, at least 58% of the ratings received for the measure *result-orientation* [TC.P.3.2] indicate it to be highly relevant for representing team capability.

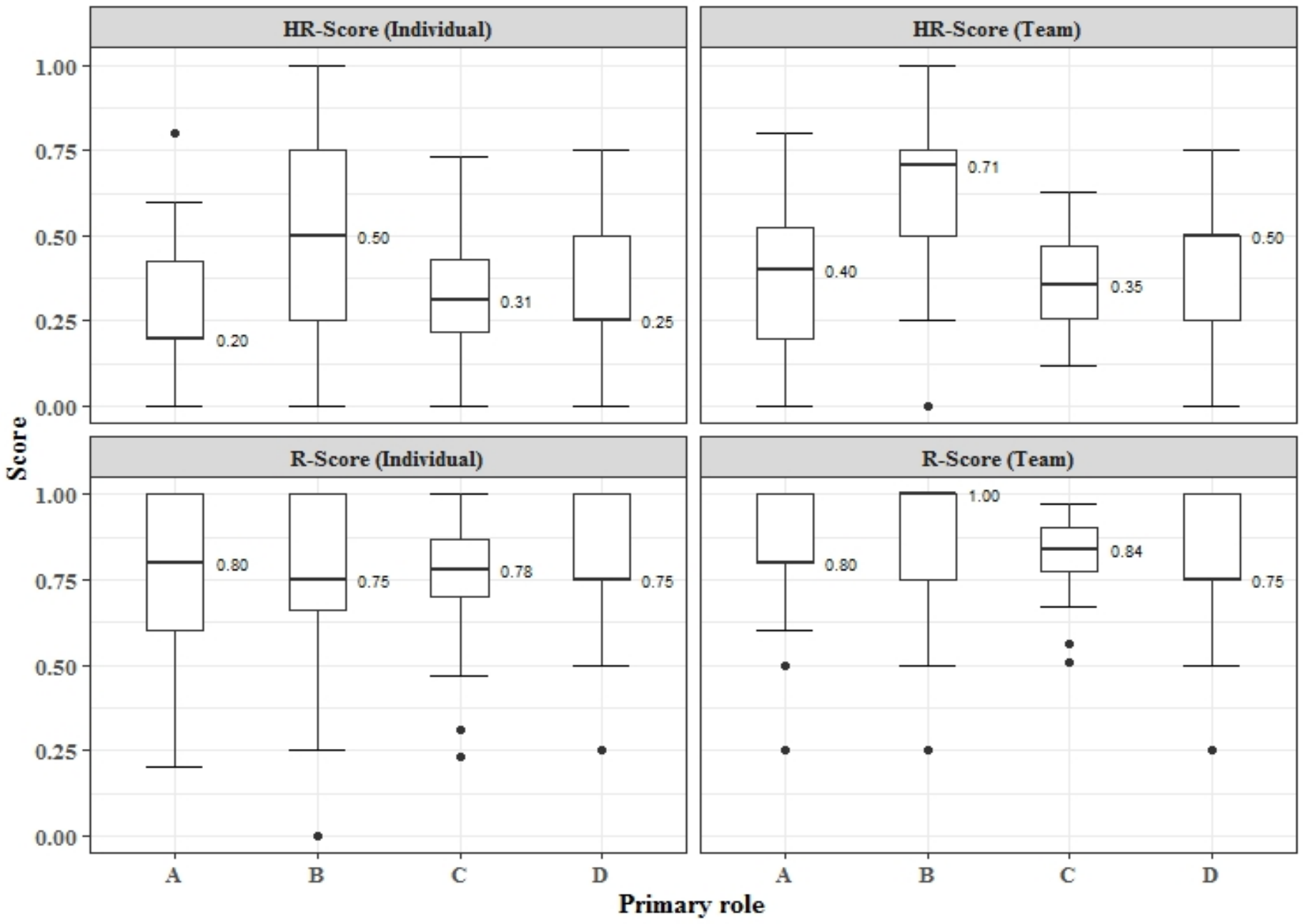

**Figure 7. Distribution of scores for capability measures;**
**Grouping criterion: Primary role**
**(A: CI engineer, B: Scrum master, C: Developer & D: Tester)**

**Table 16. Top five rated individual and team capability measures; Grouping criterion: Primary role**

| Type | CI engineer | | | Scrum master | | | Developer | | | Tester | | |
|---|---|---|---|---|---|---|---|---|---|---|---|---|
| | ID | HR-score | R-score | ID | HR-score | R-score | ID | HR-score | R-score | ID | HR-score | R-score |
| Individual capability measures | IC.S.1.9 | 0.8 | 1 | IC.S.1.9 | 1 | 1 | IC.S.2.2 | 0.73 | 0.95 | IC.P.4.7 | 0.75 | 1 |
| | IC.S.5.3 | 0.8 | 1 | IC.S.3.4 | 1 | 1 | IC.S.5.5 | 0.7 | 1 | IC.P.4.8 | 0.75 | 1 |
| | IC.S.5.5 | 0.8 | 1 | IC.S.4.16 | 1 | 1 | IC.S.3.5 | 0.68 | 0.95 | IC.S.1.5 | 0.75 | 1 |
| | IC.I.2.2 | 0.8 | 1 | IC.S.5.4 | 1 | 1 | IC.S.5.3 | 0.67 | 0.95 | IC.S.2.3 | 0.75 | 1 |
| | IC.P.1.2 | 0.6 | 1 | IC.S.5.5 | 1 | 1 | IC.S.2.3 | 0.65 | 1 | IC.S.4.1 | 0.75 | 1 |
| Team capability measures | TC.S.1.3 | 0.8 | 1 | TC.P.1.4 | 1 | 1 | TC.S.1.6 | 0.63 | 0.97 | TC.I.1.2 | 0.75 | 1 |
| | TC.P.3.3 | 0.6 | 1 | TC.P.3.1 | 1 | 1 | TC.P.5.1 | 0.63 | 0.97 | TC.P.5.2 | 0.75 | 0.75 |
| | TC.S.1.1 | 0.6 | 1 | TC.P.3.2 | 1 | 1 | TC.P.6.1 | 0.6 | 0.87 | TC.P.3.1 | 0.5 | 1 |
| | TC.S.1.2 | 0.6 | 1 | TC.P.5.2 | 1 | 1 | TC.P.3.2 | 0.58 | 0.94 | TC.P.6.1 | 0.5 | 1 |
| | TC.S.1.6 | 0.6 | 1 | TC.P.1.5 | 0.75 | 1 | TC.S.1.5 | 0.51 | 0.87 | TC.S.1.1 | 0.5 | 1 |

#### 4.2.2.4. Grouping criterion 4: Respondents' work experience

Among the five sub-groups created based on the respondents' work experience, the set of HR-scores and R-scores calculated for each capability measure were used to generate boxplots (see Figure 8). In the case of the boxplots for HR-scores of [3, 4) years sub-group (individual capability measures) and [6, 7) years sub-group (team capability measures), the large size of box and whiskers indicate that there were wide disparities among the perceptions of members. Upon comparing the distribution of HR-scores and R-scores across different sub-groups, the median for [6, 7) years sub-group can be seen to be relatively higher, indicating that the members from this sub-group perceived greater proportion of the capability measures as relevant.

Within the [6, 7) years sub-group, we can clearly see that the R-score was 1 for almost 50% of the individual and team capability measures. This informs us that all the members within [6, 7) years sub-group considered around half of the capability measures as relevant. In the case of team capability measures, the size of the boxplot for [5, 6) years sub-group appears to be smaller than the plots for rest of the sub-groups. This indicates that, among the members having [5, 6) years of work experience, there is high level of conformity among the perceptions of team capability measures.

The top measures from each sub-group are presented in Table 17 and the key aspects are discussed as follows:

- Within each of the five sub-groups, at least 66% of the ratings received for ‘responsibility’ (IC.S.5.5) indicate it to be highly relevant for representing individual capability.
- At least 61% of the ratings from [3, 4) years and [4, 5) years sub-groups indicate the following measures to be highly relevant for representing individual capability:
  - o Work ethics (*motivation to work* [IC.S.5.3])
  - o Communication (*listening skills* [IC.S.2.2])
- Within each of the five sub-groups, at least 52% of the ratings received for *cooperation* [TC.S.1.6] indicate it to be highly relevant for representing team capability.

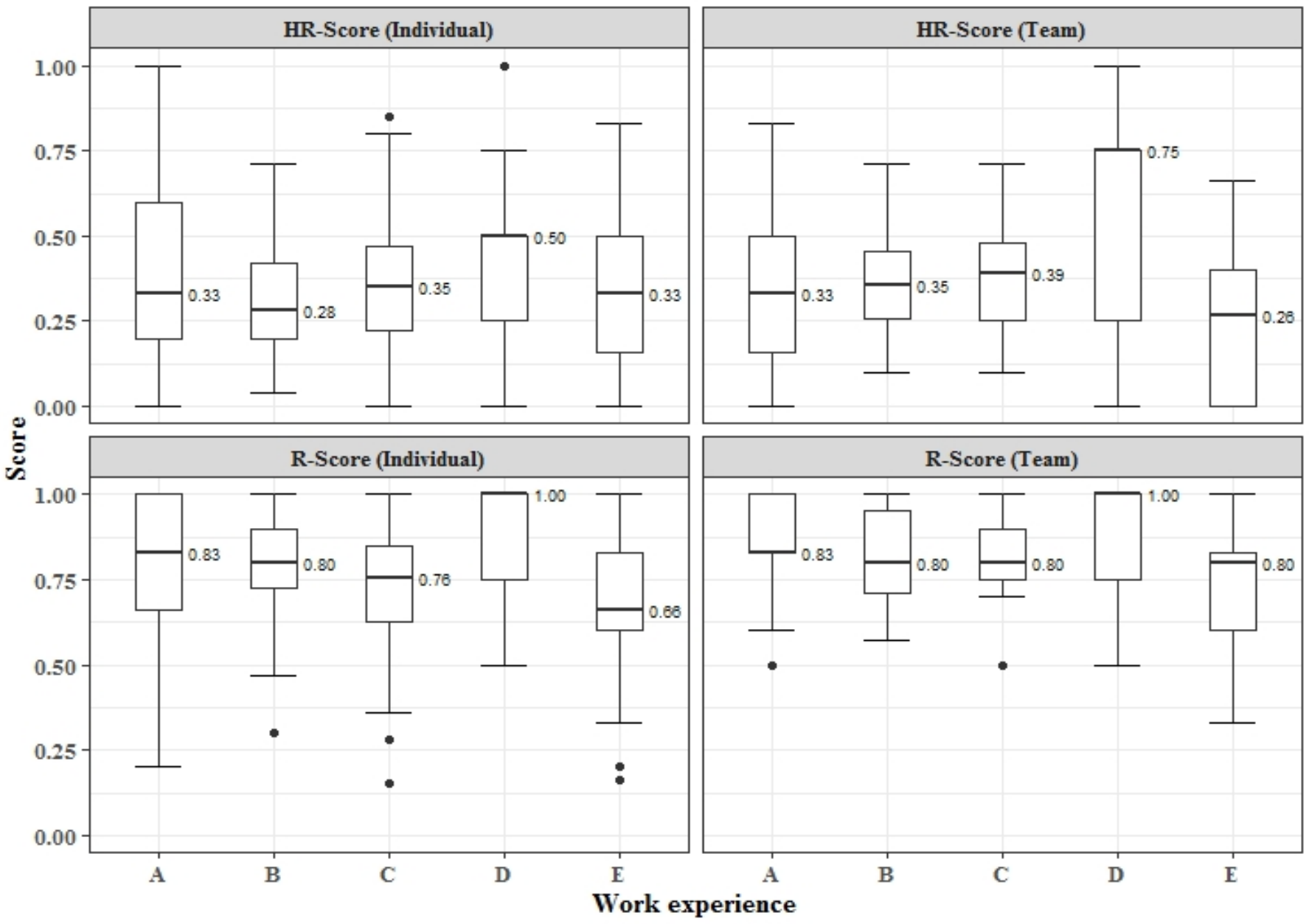

**Figure 8. Distribution of scores for capability measures;**
**Grouping criterion: Work experience**
**(A: [3, 4) years, B: [4, 5) years, C: [5, 6) years, D: [6, 7) years & E: > 7 years)**

**Table 17. Top five rated individual and team capability measures; Grouping criterion: Work experience. The IDs of measures that are common across all the sub-groups are highlighted in bold**

| Type | [3, 4) years | | | [4, 5) years | | | [5, 6) years | | | [6, 7) years | | | > 7 years | | |
|---|---|---|---|---|---|---|---|---|---|---|---|---|---|---|---|
| | ID | HR-score | R-score | ID | HR-score | R-score | ID | HR-score | R-score | ID | HR-score | R-score | ID | HR-score | R-score |
| Individual capability measures | IC.S.2.2 | 1 | 1 | **IC.S.5.5** | 0.71 | 1 | IC.S.4.20 | 0.85 | 0.95 | IC.P.10.5 | 1 | 1 | IC.S.5.3 | 0.83 | 1 |
| | IC.S.5.2 | 1 | 1 | IC.S.2.3 | 0.61 | 1 | IC.S.4.22 | 0.85 | 0.95 | **IC.S.5.5** | 1 | 1 | IC.S.5.6 | 0.83 | 1 |
| | IC.S.5.3 | 1 | 1 | IC.S.3.5 | 0.61 | 1 | IC.S.2.2 | 0.8 | 0.95 | IC.I.1.4 | 1 | 1 | IC.P.1.2 | 0.66 | 1 |
| | IC.S.5.4 | 1 | 1 | IC.S.2.2 | 0.61 | 0.95 | IC.S.4.1 | 0.8 | 0.85 | IC.P.5.1 | 0.75 | 1 | IC.P.1.3 | 0.66 | 1 |
| | **IC.S.5.5** | 1 | 1 | IC.S.5.3 | 0.61 | 0.95 | **IC.S.5.5** | 0.76 | 1 | IC.P.6.3 | 0.75 | 1 | **IC.S.5.5** | 0.66 | 1 |

| | **[3, 4) years** | | | **[4, 5) years** | | | **[5, 6) years** | | | **[6, 7) years** | | | **> 7 years** | | |
|---|---|---|---|---|---|---|---|---|---|---|---|---|---|---|---|
| | **ID** | **HR-score** | **R-score** | **ID** | **HR-score** | **R-Score** | **ID** | **HR-Score** | **R-Score** | **ID** | **HR-Score** | **R-Score** | **ID** | **HR-Score** | **R-Score** |
| Team capability measures | TC.P.1.2 | 0.83 | 1 | TC.P.5.1 | 0.71 | 0.95 | TC.P.5.1 | 0.71 | 0.9 | TC.P.3.2 | 1 | 1 | **TC.S.1.6** | 0.66 | 0.83 |
| | TC.S.1.2 | 0.66 | 1 | TC.P.3.1 | 0.57 | 1 | TC.P.3.2 | 0.66 | 0.8 | TC.P.4.1 | 1 | 1 | TC.P.2.3 | 0.6 | 0.8 |
| | TC.S.1.3 | 0.66 | 1 | TC.P.6.1 | 0.57 | 1 | TC.P.5.2 | 0.57 | 1 | TC.P.5.2 | 1 | 1 | TC.P.5.1 | 0.5 | 1 |
| | **TC.S.1.6** | 0.66 | 1 | TC.P.5.2 | 0.57 | 0.8 | **TC.S.1.6** | 0.57 | 0.95 | TC.S.1.5 | 1 | 1 | TC.S.1.1 | 0.5 | 1 |
| | TC.P.3.2 | 0.66 | 0.83 | **TC.S.1.6** | 0.52 | 0.95 | TC.S.1.5 | 0.57 | 0.85 | **TC.S.1.6** | 1 | 1 | TC.S.1.5 | 0.5 | 1 |

#### 4.2.2.5. Grouping criterion 5: Respondents' team size

Among the three sub-groups created based on the team size of respondents, the set of HR-scores and R-scores calculated for each capability measure were used to generate boxplots (see Figure 9). Upon inspecting the boxplots for HR-scores, it is evident that the members whose team size ranged from 11 to 15 perceived greater proportion of capability measures as highly relevant. However, the elongated whiskers in the case of this sub-group indicate that there were greater dispersions in the perceptions of members. In the case of [1, 5] and [6, 10] sub-groups, the R-scores for all the team capability measures were above 0.50. This informs us that, within these sub-groups, all the team capability measures were regarded as relevant by at least 50% of the members. Here, the small sized boxplots indicate higher conformity among the perceptions of members.

The list of top measures from each group is presented in Table 18 and the key aspects are discussed as follows:

- Within each of the three sub-groups, at least 68% of the ratings received for *responsibility* [IC.S.5.5] indicate it to be highly relevant for representing individual capability.
- In relation to individual capability measures, teams with [1, 5] members and [6, 10] members commonly regarded *desire to improve things* [IC.S.4.20] to be highly relevant. Whereas teams with [6, 10] members and [11, 15] members commonly considered the measure *listening skills* [IC.S.2.2] to be highly relevant.
- Among the three sub-groups, at least 50% of the ratings received for *result-orientation* [TC.P.3.2] indicate it to be highly relevant for representing individual capability.
- At least 56% of the ratings from teams with [1, 5] members and [6, 10] members indicate the following measures to be highly relevant for representing team capability:
    - o Growth (*active learning and improvement* [TC.P.5.1])
    - o *Cooperation* [TC.S.1.6]
- Teams with [6, 10] members and [11, 15] members commonly considered *advancement* [TC.P.5.2] to be highly relevant for representing team capability.

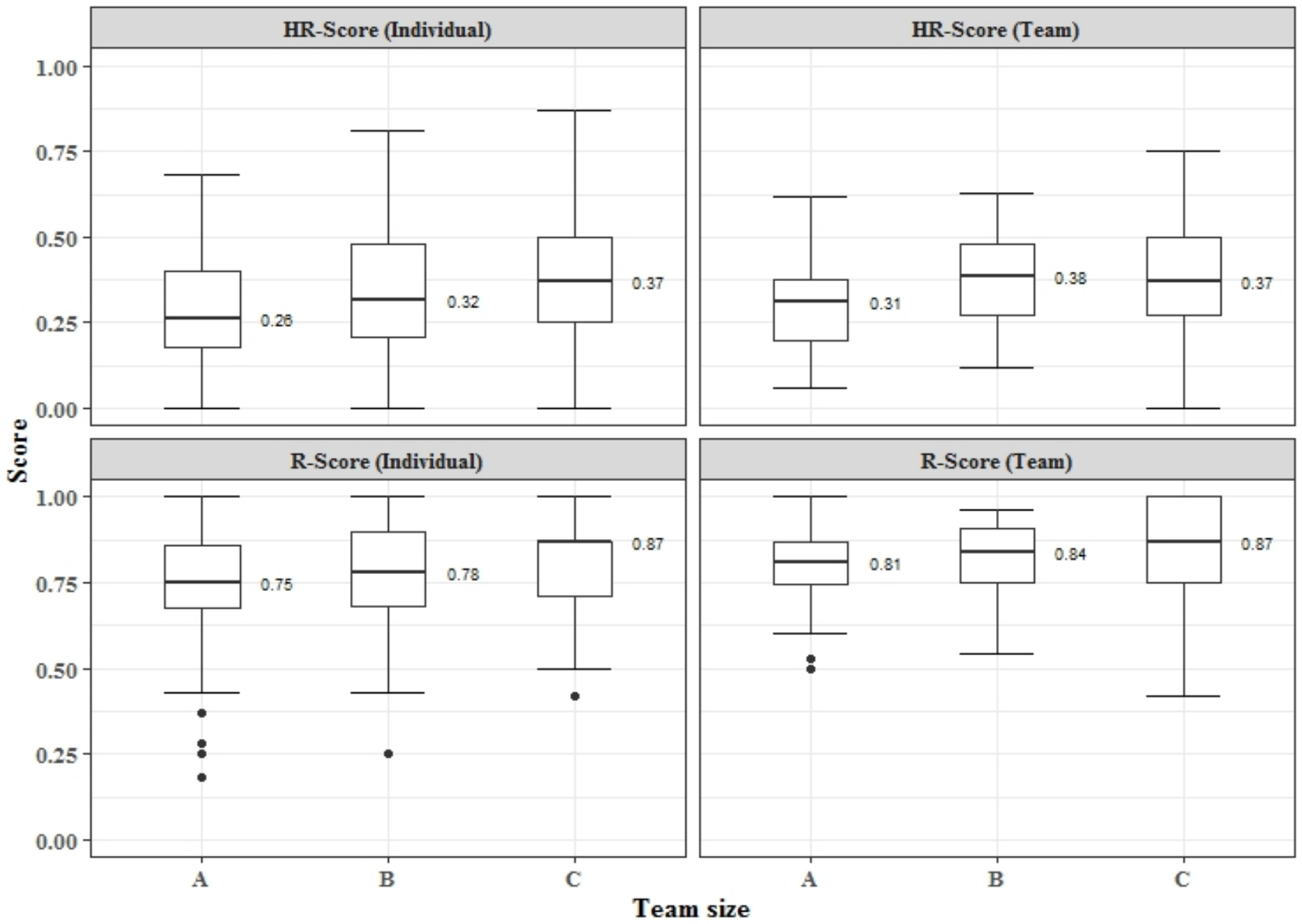

**Figure 9.Distribution of scores for capability measures;**
**Grouping criterion: Team size**
**(A: [1, 5] members, B: [6, 10] members & C: [11, 15] members)**

**Table 18. Top five rated individual and team capability measures; Grouping criterion: Team size. The IDs of measures that are common across all the sub-groups are highlighted in bold**

| Type | [1, 5] members | | | [6, 10] members | | | [11, 15] members | | |
|---|---|---|---|---|---|---|---|---|---|
| | ID | HR-score | R-score | ID | HR-score | R-score | ID | HR-score | R-score |
| Individual capability measures | **IC.S.5.5** | 0.68 | 1 | IC.S.2.2 | 0.81 | 0.96 | IC.S.3.4 | 0.87 | 1 |
| | IC.S.3.5 | 0.68 | 0.93 | IC.S.5.3 | 0.78 | 0.96 | IC.S.4.20 | 0.75 | 1 |
| | IC.S.4.20 | 0.68 | 0.87 | IC.S.2.3 | 0.75 | 1 | IC.S.4.21 | 0.75 | 1 |
| | IC.S.1.9 | 0.62 | 1 | **IC.S.5.5** | 0.75 | 1 | **IC.S.5.5** | 0.75 | 1 |
| | IC.I.2.1 | 0.62 | 0.93 | IC.S.5.4 | 0.69 | 1 | IC.I.2.2 | 0.75 | 1 |
| Team capability measures | TC.P.5.1 | 0.62 | 0.87 | TC.S.1.6 | 0.63 | 0.93 | TC.P.5.2 | 0.75 | 1 |
| | TC.S.1.6 | 0.56 | 0.93 | TC.P.5.1 | 0.57 | 0.96 | **TC.P.3.2** | 0.75 | 0.87 |
| | TC.P.6.1 | 0.56 | 0.81 | TC.S.1.2 | 0.57 | 0.93 | TC.P.2.2 | 0.62 | 0.87 |
| | TC.S.1.5 | 0.56 | 0.75 | **TC.P.3.2** | 0.54 | 0.93 | TC.I.1.2 | 0.62 | 0.87 |
| | **TC.P.3.2** | 0.5 | 0.93 | TC.P.5.2 | 0.54 | 0.9 | TC.I.1.1 | 0.62 | 0.75 |

#### 4.2.2.6. Key insights from the sub-groups

We inspected and compared the tables in this section 4.2.2 (Table 14, Table 15, Table 16, Table 17 and Table 18) to find highly regarded capability measures that were common across sub-groups. We observed the individual capability measure *responsibility* to be common among the sub-groups created based on organizational domain, work experience and team size. Whereas among team capability measures, we observed *cooperation* to be common among all the sub-groups created based on ASD method and work experience. These findings were observed to be in line with the highly regarded capability measures from the overall sample.

Most members from various sub-groups perceived the measures as relevant for representing individual and team capability. However, due to the small and unequal group sizes in our study, we cannot firmly declare the top capability measures for each group, based on the observed HR-scores or R-scores.

## 4.3. Identifying additional capability measures (RQ.3)

In total, 10 respondents answered the two open-ended questions from the last section of the questionnaire and among them, eight respondents indicated additional measures in relation to individual capability and three respondents indicated measures in relation to team capability. The responses provided to the two open-ended questions were straight forward i.e., responses consisted of the names/very brief description of capability measures. Therefore, it was not applicable to perform coding or qualitative analysis for those responses. The open-ended data (free text) can be found in the Appendix C of the supplementary material. We hardly received any qualitative evidence, and whatever evidence was provided, it was meant to complement the set of capabilities that had already been selected by the respondent, and not to validate those capabilities.

We observed that the context of some of the measures indicated by the respondents corresponded to the measures already existing in the questionnaire. In the light of our questionnaire presenting more than 160 capability measures, we believe it would be difficult for respondents to retain all the measures in short-term memory. So, upon comparing the responses of open-ended questions with the existing catalogue of measures from our SLR, we identified seven additional individual capability measures and one team capability measure (see Table 19) that were not discussed within the context of capability measurement by former SE studies. Based on the context of the measures, we have further classified them into pertinent sub-categories as shown in Table 19.

In relation to individual capability measures, the following measures were indicated by no more than one respondent: the quality of relying on one's resources without needing support from other people (self-reliance), proficiency in local language to engage in informal conversations with others, the ability to prioritize when multiple tasks are allocated, proficiency in reviewing code written by others and the pace at which a person works. Among these, proficiency in code reviewing and communicating in local language were classified under professional category, self-reliance and task-prioritization were classified under social category.

Additionally, two respondents indicated the quality of being able to adjust to new or changing technologies as a measure of individual capability. Although this measure appears to be identical to *flexibility* [IC.S.5.1], we believe being able to adapt is associated with long-term changes, as opposed to being flexible, which is associated with more short-term alterations. Further, a person's ability to optimize existing solutions/systems was indicated by two respondents. This seemed to be an innovative capability measure that is analogue to the measure 'desire to improve things' (IC.S.4.20). Finally, in relation to team capability measures, two respondents indicated a team's ability to make decisions as a measure relevant for representing the team's capability. This measure has been categorized under professional category after comparing with existing catalogue of team capability measures.

**Table 19. Additional capability measures identified from open-ended questions**

| Individual capability measures | | | Team capability measures | | |
|---|---|---|---|---|---|
| **Sub-category** | **Capability measure** | **Count** | **Sub-category** | **Capability measure** | **Count** |
| Software Construction (IC.P.4) | Proficiency in code reviewing | 1 | Miscellaneous (TC.P.6) | Ability to make decisions | 2 |
| Miscellaneous (IC.P.13) | Proficiency in local language for communication | 1 | | | |
| | Work pace | 1 | | | |
| Affective (IC.S.1) | Self-reliance | 1 | | | |
| Work ethics (IC.S.5) | Task prioritization | 1 | | | |
| | Ability to adapt to changing technologies | 2 | | | |
| Managing change (IC.I.5) | Ability to optimize existing solutions/systems | 2 | | | |

# 5. Discussion

In this section, we discuss how the main findings from our research questions relate to former SE studies. Further, we discuss the limitations of our study, implications of our findings for research and industrial practice, and future work.

## 5.1. Comparison with related work

In our study, the response rate obtained for the original sampling frame (17%) was observed to be in line with other surveys that targeted at gathering agile practitioners' perceptions (e.g., [6], [46], [93]). In a field such as SE, where the sampling frame is likely to be small [49], we believe the response rate of 17% is fair, especially in the light of the length of our questionnaire and the relatively small sampling frame employed. The respondents in our study were not only experienced professionals, but also were associated with diverse roles and domains. This makes our results applicable to a wider audience.

By inspecting columns from Table 4 through Table 8, we noticed that, except for five measures, other individual and team capability measure was indicated as relevant by more than 50% of the respondents. Thus, we observed the measures to be prevalent among agile practitioners and the results from our survey bestows greater confidence in the pertinence of our SLR [22] findings to the area of capability measurement in ASD. We further believe that using multiple sources of information (literature and agile practitioners) and multiple research methods (SLR and survey) to understand a phenomenon (capability measurement) leads to more accurate results and conclusions.

Four individual capability measures that failed to earn the 'approval' of more than 50% of the respondents were: *years in company*, *mixes personal and work goals*, *introversion* and *feeling*. Colomo-Palacios et al. [85] considered *years in a company* and *total work experience* of a person as factors that contribute towards forming the most suitable team for a work package. However, our results indicate that practitioners perceived the number of years spent in a company as trivial in relation to an agile team member's capability. For this reason, we have not categorized the sample of respondents on the basis of years in a company while answering RQ.3 and RQ.4.

Besides *years in a company*, the majority of the respondents also regarded the measure *mixes personal and work goals* as trivial. Although a person's quality of associating personal goals and work goals was reported in a competency model as a measure of individual capability [11], we think that, in our study, some of the respondents might have perhaps foreseen the risk of neglecting work goals under the circumstances where the motive of personal gain overrides the achievement of work objectives.

The other two measures considered trivial pertain to an individual's personality. We observed that practitioners' impression about *feeling* personality dimension was in line with the findings from former SE studies. A study that investigated the differences in personality types among software developers who used agile and non-agile methodologies, reported *feeling* personality dimension to be more dominant among developers practicing non-agile methodologies [20]. Additionally, another study reported that people with *thinking* personality type can write more efficient code than people with *feeling* personality type [94]. The view of respondents regarding *introversion* does not come as a surprise as introverts tend to communicate within smaller groups and often respond to conversation rather than staring it [95]. Such characteristics would obviously be less favored within agile teams which require more social interactions [96].

Interestingly, in the case of extraversion, we observed an equal split of responses within the sample. Upon reviewing former SE literature, we noticed that while extroverted professionals were credited for being communicative, making *extroversion* a preferable personality type for agile teams, extrovert professionals on the other hand, were also reported to be relatively impatient on complex software development tasks where they expressed themselves to be intolerant to slow project velocities [96]. This could perhaps be a plausible explanation for the differences among practitioners' perceptions on extraversion dimension.

Among all the capability measures, *responsibility* and *questioning skills* were indicated by the whole sample to be relevant for representing individual capability. In specific, *responsibility* was commonly regarded as highly relevant across the sub-groups created based on organizational domain, work experience and team size. The reason for this could be attributed to the principles of agility, which promote shared responsibility and self-management among team members. Agile practices encourage voluntary and proactive participation of all team members [97], [98]. In specific, self-management is a defining characteristic for Scrum methodology where a team is accorded full authority for deciding about ways to achieve goals [85], [99]. Since the majority (68%) of the respondents in our study adopted Scrum practices in their work, the significant authority and responsibility delegated to them would have perhaps influenced their perception of responsibility and questioning skills.

On the other hand, the team level aspects: *cooperation*, *active learning and improvement*, *communication skills*, *programming language experience* and *creative exploration and exploitation*, were observed to be top rated measures for representing team capability. Especially, *cooperation* was commonly reflected as highly relevant among all the sub-groups created based on ASD method and work experience. These findings were in accordance with former SE studies where researchers emphasized on the importance of team level *cooperation* [24], [88] and *programming language experience* [44], [45]. Besides that, the aforementioned five factors have been reported to affect the performance [38], [88] and productivity [44], [24] of an agile team and this could perhaps be the reason why the measures were perceived to be highly crucial for determining team capability.

When looking from the practitioners' perception of capability measures, we observed that the top measures perceived as highly relevant for characterizing capability pertained to social category. Based on FSC-2 ranks, the top five ranked individual capability measures (*responsibility*, *listening skills*, *questioning skills*, *team participation skills* and *being teamwork oriented*) belonged to the social category, where at least 63% of the respondents considered them as very crucial. On the other hand, two out of top five team capability measures (*cooperation* and *communication skills*) belonged to the social category, where at least 48% of the respondents considered them as highly relevant. The dominance of social aspects in characterizing capability was seen to be in line with the findings from our SLR [22], where 75% of the primary studies reported measures in relation to social aspects of individuals or teams.

The reason for the dominance of social aspects could be attributed to one of the core values of the agile manifesto, which is "individuals and interactions over processes and tools" [87]. ASD is a sociotechnical practice and its processes are designed to capitalize on each individual and team's unique strengths [100], [101]. Former SE literature reported social aspects like soft skills and non-technical skills to be influencing software quality [102], team performance [19] and productivity [44]. These skills, when combined adeptly, have been reported to maximize work effectiveness and were opined to be even more important than the traditional qualifications and technical skills for personal and professional success [40].

By grouping respondents based on demographics, we explored whether there were significant differences between the perceptions of sub-groups. In cases where we observed a difference, the effect size was calculated to determine how likely such differences can also be observed in other studies. While aspects like ASD methodology, primary role and work experience had a relatively strong effect on how respondents perceived some capability measures, we noticed that the work experience of respondents had a strong effect on the perceptions of the majority of the capability measures (four individual capability measures and one team capability measure). A former study that investigated the effect of work experience on the application of agile methods in organizations [103], uncovered that use of agile practices vary with experience of practitioners. While less experienced practitioners were observed to use few technical practices, practitioners with more experience adopted a wide range of practices. We believe the disparities in exposure to technical practices could perhaps be an underlying factor influencing the perceptions of capability measures.

## 5.2. Limitations

Our questionnaire included two open-ended questions for adding additional capability measures. But these questions were not marked as mandatory. Further, our questionnaire did not include any follow-up questions requesting subjects to elaborate the reasons why they perceived a capability measure as irrelevant. These aspects can be seen as limitations of our questionnaire, however, an elaborate explanation for such open questions would require a significant time commitment by subjects [76] and this could affect the number of participants. Further, as we wanted to adhere to the best practices suggested by King et al. [71], on examining the duration of our pilot session, we decided not to include any other questions.

In Section 4.2.1, for analyzing the differences in the ratings of the capability measures across different sub-groups, we used the K-W test. As the K-W test requires sample size of each group to be at least five [89], we have conducted analysis only when group size was greater than five and have omitted the sub-groups with too few observations (less than five) from this test. This is one of the limitations of this study.

Another limitation is in relation to reporting results of RQ.2.2. Although the scoring criteria discussed in Section 4.2.2 includes four different scores (HR, R, T and HT scores), while analyzing perceptions across sub-groups, we only presented an analysis for HR and R scores and the boxplots were also presented only for these two scores. This was because of space constraints. Due to the same reason, in that section, we have also limited our discussion on scores to the top five capability measures from each sub-group.

## 5.3. Implications and lessons learned

In practice, we believe that our study could be useful to managers while assembling teams and while selecting teams to a project. Managers can use the capability measures vetted by agile practitioners towards deriving a tailored list of measures that focus on industrial needs. For example, managers can maintain separate checklists for the role of tester and developer and use them towards recruiting new members. Further, to suit the requirements of a customer or a project, managers can generate a checklist to assemble a team with all the relevant skills. To identify definitive measures for a checklist, managers can either refer to the heatmaps included in Section 4.1, or survey a specific group of practitioners, as done in this study, and use the scoring criteria discussed in Section 4.2.2 for identifying top measures.

We further suggest software organizations to maintain a competence repository of their employees and to vividly incorporate capability assessments in their decision-making process. This could perhaps be facilitated by employing a capability-centric agile support tool [29]. Practitioners can add the top 22 individual capability measures (Section 4.1.1.4) and 5 team capability measures (Section 4.1.1.4) identified in this study as assessment parameters in the agile support tool. These parameters aid in quantifying an individual's or team's skills and in turn help in differentiating precise personnel. Adopting such an agile support tool not only coordinates tasks like assessment of capability measures, team composition and task allocation, but also benefits an organization in diagnosing beyond the project planning problems. The tool lets organizations take decisions about offering training to their employees, in order to improve their skills.

The strong effects of criteria such as organizational domain, ASD methodology, primary role and work experience of agile practitioners observed in our analysis are highly likely to be observed in other investigations within an ASD context. However, the small sample sizes of sub-groups used in our analysis do not permit us to say anything definitive about trends. In order to uncover deeper insights and help in generalizing the findings, researchers are encouraged to formulate hypotheses using the measures that were perceived differently across sub-groups (Section 4.2.1). These hypotheses can then be tested by recruiting a sample whose size is larger than the one in our study.

As we could not find an obvious forum for contacting a large number of agile practitioners for our survey, apart from approaching personal contacts, we also resorted to using LinkedIn for recruiting subjects. Since there are currently no clear and established standards for identifying survey subjects over social media [70], [71], we used a search-string-based strategy over LinkedIn to recruit survey subjects. Besides showing direct connections, interestingly, the search also retrieved a list of secondary connections possessing the relevant skills. Based on our experience of using this strategy, we learned that collaborating with people who have industry and academia connections on LinkedIn can aid in identifying diverse subjects and in turn help towards recruiting a bigger sample.

One of the key novelties in our study's approach towards informing practice is to blend the perspectives from scientific literature and industrial practice. We learned that this approach is effective as the results from the two perspectives are complementary to each other. In this regard, we recommend researchers who aim to inform practice in their forthcoming studies, to first synthesize the state of the art and subsequently use those findings towards determining the state of the practice. Next, a union of the results from both perspectives can be derived and ultimately be used to advocate for practice.

## 5.4. Threats to validity

This section discusses some of the threats that could affect the validity of our results.

**Construct validity:** This threat relates to the issues that might arise because of improper design of the survey instrument. In order to ensure that the instrument properly measures what it is supposed to measure, we requested two experienced external reviewers who had experience with ASD to assess our questionnaire. Their suggestions regarding presentation and clarity were duly addressed. We were unable to identify any survey similar to ours that could have been employed to help assess our instrument for criterion validity [51]. We believe the threat related to construct validity has been mitigated as our instrument was iteratively designed and updated based on the results from our SLR [22].

Moreover, we used an empirically evaluated checklist [54] to guide our survey design and to audit our survey report. Each checklist item was scored based on the corresponding reflections in our survey report. The scoring scheme used by Molléri et al. [54] was adopted for our evaluation, where each checklist item was either ranked as fully addressed (1 point), partially addressed (0.5 point), not addressed (0 point) or not applicable (NA). Besides scoring, we also reported the reasoning to score each item, as suggested for checklist usage. In our evaluation, the resulting score was 34 out of 38 (i.e., 34 items were fully addressed). The remaining four items were scored as NA. The details of our checklist evaluation can be found in Appendix D of the supplementary material. In addition to fully addressing most of the items on the checklist, the main author of this checklist, Dr. Molléri, was one of the two reviewers who piloted our survey. Therefore, we have confidence in the validity of the survey instrument.

**Internal validity:** This threat relates to the issues with confounding factors or irrelevant respondents that could potentially introduce a systematic error or bias in the study results. The following steps were taken to mitigate this threat: (1) The survey homepage as well as invitation explicitly mentioned that the survey was intended for seeking the opinions of practitioners with actual experience of working in agile teams. Besides asking respondents about their experience, we also inquired about their team size and ASD methods practiced in their team. All the 60 respondents answered these questions. (2) In order to minimize the evaluation apprehension of respondents, they were explicitly informed that there were no right, or wrong answers and were also assured of their anonymity.

**External validity:** This threat relates to the generalizability of the study findings. Since we recruited subjects based on convenience sampling, our results would only be applicable to those agile teams and organizational domains that share similar characteristics to our sample of respondents. However, steps such as employing social-media and a snowball approach were adopted to acquire a broader representative sample. We believe these steps aided in obtaining a sample that was quite heterogeneous in terms of experience and job role.

Upon analyzing the perceptions among the sample of respondents, for the majority of the capability measures from Table 4 through Table 8, we observed the lowest percentage of ratings in the 'irrelevant' column. However, we should probably not infer much from those small numbers because doing so presents a very strong statement [49].

Further, as some grouping criteria were observed to have a medium effect size on the perceptions of respondents, there is an indication that the differences observed in our findings would not occur in some agile teams under similar context. However, expanding this research by recruiting a bigger sample could help in uncovering differences that can be representative for agile teams.

**Conclusion validity:** This threat relates to the possibility of arriving at incorrect conclusions due to errors emanating from inadequate statistical tests. In order to point out potential areas for future research, in this study, we majorly used frequencies and percentages to analyze respondents' perceptions. Moreover, as K-W test requires each group under consideration to have at least five observations [89], while studying the differences between sub-groups in our study, only the groups whose size was greater than or equal to five were considered for our analysis.

### 5.5. Future work

The capabilities of software professionals influence team climate [28], team performance [19], [18], [13] and also affect software quality [104], [105]. So, a promising and relevant research direction for future work would be to see how the individual and team capability measures could be used as predictors for forecasting team climate, performance, and product quality. Such an investigation not only aids in gathering empirical evidence but also helps in exploring the causal relations, with respect to the impact of capability measures on the aforementioned organizational outcomes.

## 6. Conclusion

This paper presents the results of an empirical study that was executed to understand which individual and team level measures would be appropriate for characterizing the capability of an agile team and its members. The study employed a survey procedure and used an online questionnaire as instrument for collecting responses from 60 agile practitioners across six countries.

Upon analyzing the survey responses, non-technical skills such as *responsibility*, *listening skills* and *questioning skills* emerged as the top capability measures, which were indicated by at least 63% of respondents as highly relevant for characterizing the capability of an agile team member. Whereas on team level, *cooperation*, *active learning and improvement* and *communication skills* were observed to be the top capability measures which were indicated by at least 48% of the respondents as highly relevant for characterizing the capability of an agile team. Among all the capability measures, respondents unanimously indicated *questioning skills* and *responsibility* to be relevant for representing the capability of an individual.

We observed that respondents' work experience had a significant relatively strong effect on how they perceived five capability measures. Similarly, criteria like the agile methodology used at respondents' organization and respondents' primary role, also had a significant relatively strong effect on their views of capability measures.

Upon comparing the top ranked capability measures across different sub-groups, we observed that the majority of the sub-groups considered *responsibility* to be a highly relevant measure for characterizing the capability of an agile team member. On the other hand, *cooperation* was regarded by majority of the sub-groups as a highly relevant measure for characterizing the capability of an agile team.

We believe exploring practitioners' perceptions on individual and team capability measures, enables the use of the measures towards improving team formation in ASD. Some of the capability measures additionally recognized from this survey, such as *work pace, self-reliance*, *task prioritization*, differ from those identified by our previous systematic literature review. This suggests that the use of different research methods and contexts may lead us to a better realization of a given phenomenon of interest.

## Supplementary material

Supplementary material for this paper is available at: http://urn.kb.se/resolve?urn=urn:nbn:se:bth-18762

## Acknowledgement

We would like to thank all those who participated in our survey. This research was partially funded by the KK-stiftelsen-HÖG AgileSec research project.